\documentclass[draft]{agujournal2019}
\usepackage{url} 
\usepackage{lineno}
\usepackage[inline]{trackchanges} 
\usepackage{soul}

\usepackage{graphicx}
\usepackage{amssymb}
\usepackage{float}
\usepackage{bm}
\usepackage{array}
\usepackage{amsmath}
\usepackage{booktabs} 
\usepackage{multicol}

\linenumbers
\graphicspath{{Pic/}}

\draftfalse

\journalname{}

\begin{document}

%
%


\title{Asymmetry in earthquake interevent time intervals}

%
%




\authors{Yongwen Zhang\affil{1,2,3}, Yosef Ashkenazy\affil{2} \& Shlomo Havlin\affil{3}}

\affiliation{1}{Data Science Research Center, Faculty of Science, Kunming University of Science and Technology, Kunming 650500, Yunnan, China,}
\affiliation{2}{Department of Solar Energy and Environmental Physics, The Jacob Blaustein Institutes for Desert Research, Ben-Gurion University of the Negev, Midreshet Ben-Gurion 84990, Israel,}
\affiliation{3}{Department of Physics, Bar-Ilan University, Ramat Gan 52900, Israel}





\correspondingauthor{Yongwen Zhang}{zhangyongwen77@gmail.com}




\begin{keypoints}
\item We study the asymmetry of earthquake interevent time intervals which exhibits a crossover.
\item We suggest that the mechanism of the observed asymmetry is related to the earthquake triggering processes.
\item The observed asymmetry is better reproduced by an improved ETAS model developed recently.
\end{keypoints}

%
%

%
%


\begin{abstract}
Here we focus on a basic statistical measure of earthquake catalogs that has not been studied before, the asymmetry of interevent time series (e.g., reflecting the tendency to have more aftershocks than spontaneous earthquakes). We define the asymmetry metric as the ratio between the number of positive interevent time increments minus negative increments and the total (positive plus negative) number of increments. Such asymmetry commonly exists in time series data for non-linear geophysical systems like river flow which decays slowly and increases rapidly. We find that earthquake interevent time series are significantly asymmetric, where the asymmetry function exhibits a significant crossover to weak asymmetry at large lag-index. We suggest that the Omori law can be associated with the large asymmetry at short time intervals below the crossover whereas overlapping aftershock sequences and the spontaneous events can be associated with a fast decay of asymmetry above the crossover. We show that the asymmetry is better reproduced by a recently modified ETAS model with two triggering processes in comparison to the standard ETAS model which only has one.
\end{abstract}

\section*{Plain Language Summary}
Earthquakes are often associated with non-equilibrium and nonlinear underlying processes which can lead to asymmetric behavior in metrics derived from earthquake records. By asymmetry we are referring to `the tendency of more events to occur after a previous one than before the next one' or vice versa. In earthquake sequences the main source of asymmetry is the occurrence of large numbers of aftershocks due to the earthquake triggering. We find here that the distributions of interevent time increments in real seismic catalogs are asymmetric and that the degree of asymmetry is characterized by a scaling function that exhibits a crossover, from a high asymmetry at short times to low asymmetry at long times. We suggest that different earthquake triggering processes are associated with these two distinct regimes of asymmetry. We apply the asymmetry analysis to an earthquake forecasting model--the Epidemic--Type Aftershock Sequence (ETAS) model and find that the new generalized ETAS model that includes both short- and long-term triggering mechanisms better reproduces the observed asymmetry than the standard ETAS model.

%
%

\section{Introduction}

Earthquakes are a major threat to society in many countries around the world. Currently, a skillful and trustworthy earthquake forecasting approach for both short and long time scales is missing. Yet, it is necessary to establish reasonable reduction strategies of seismic risk and enhance alertness and resilience. In most cases, seismologists are not yet able to predict individual large earthquakes even very close to the event \cite{jordan2011operational,DeArcangelis2016}.

Earthquake catalogs are usually restricted to specific regions and include the magnitude, location, and time of earthquakes. Several seismic laws have been discovered based on earthquake records. According to the Gutenberg-Richter law, the number of earthquakes $N$ (above a magnitude $M$) drops exponentially with the magnitude such that, $log_{10} N=a-bM$, where $b\approx 1$ and $a$ is related to the earthquake rate \cite{Gutenberg1944a}. Most earthquakes are distributed along active seismic faults which can be clearly seen in the global catalog \cite{Ide2013}. In addition, aftershocks occur around the epicenter of the mainshock and the distribution of distances from the mainshock follows a power law decay \cite{Ogata1988,Huc2003,Marsan2008}, which is related to the static or dynamic stress triggering mechanism \cite{Felzer,Lippiello2009a}.

The temporal occurrence of spontaneous earthquakes (mainshocks) are commonly assumed to follow a Poisson process with an underlying stationary rate \cite{Ogata1988}. The Omori law states that the occurrence rate of aftershocks follows as a power law decay with time \cite{Utsu1961,UTSU1972}. The probability distribution of the (scalar) interevent times of successive earthquakes in a certain region has been found to satisfy a scaling function; it is well fitted by a general gamma distribution in real data \cite{Bak2002,Corral2003,Corral2003a} similar to that found later in rock fracture experiments in laboratories \cite{Davidsen2007}. Some of the theoretical framework of the interevent times is based on the Gutenberg-Richter and the Omori laws \cite{Saichev2006,Sornette2008}. Yet, there is some criticism regarding the universal scaling with the region size \cite{Touati2009}.

Another dominant feature of earthquakes is the clustering (memory) in space and time \cite{Zaliapin2008,Yehuda2013}, generally at shorter time scales, including those for earthquake aftershock sequences and swarms. In addition, long-range memory in the time series of interevent times has been found using detrended fluctuation analysis (DFA) \cite{Lennartz2008}; strong memory was also found using the conditional probability of successive events \cite{Livina2005}. Some clustering models such as the Epidemic--Type Aftershock Sequence (ETAS) model \cite{Ogata1998} and the short-term earthquake probability (STEP) model \cite{Woessner2010} have been developed based on the short-term spatiotemporal clustering in earthquakes. In the ETAS model, the productivity parameter $\alpha$ is critical in controlling the short-term memory of interevent times \cite{Fan2018b}. Furthermore, an extended analysis of both short and long-term memory of interevent times in real data and the ETAS model \cite{Zhang2019} indicated that the inferred memory at all timescales cannot be captured by the ETAS model. A generalized (bimodal) ETAS model with two $\alpha$-values was proposed to capture short- and long-term aftershock triggering mechanisms \cite{Zhang2020}; this model reproduced the observed memory behavior in both short and long-time scales as found in real catalogs. This could be due to a sudden stress change in short-time scale and subsequent viscous relaxation in long-time scale.

The occurrence of aftershocks produces an obvious asymmetry in the time series, with more events after a previous one than before the next on the timescales of a single sequence. However, we may expect this asymmetry to degrade at longer timescales, where spontaneous events and the likelihood of overlapping aftershock sequences destroying the correlation increases, as proposed by   \citeA{Touati2009}. Asymmetry widely exists in nature \cite{An2004,Hutchinson2013} in time series for  various geophysical phenomena including the glacial-interglacial cycles (rapid warming followed by gradual cooling), the sunspot cycle (11 years) \cite{hoyt1998group}, and river flow which decays slowly and increases rapidly \cite{Livina2003a}. In many cases, such asymmetry can be related to underlying non-equilibrium and nonlinear underlying processes in a physical system \cite{King1996,Schreiber1996}. For instance, in the climate system, due to cyclone activity, surface daily mean temperature warms gradually and cools rapidly at the mid-latitudes leading to the temporal temperature asymmetry in the temperature time series \cite{Ashkenazy2008}. Here, we investigate asymmetry in earthquake time series. For triggered events, the Omori law implies that the interevent time increases with time after a mainshock. Thus, one expects asymmetry in earthquake catalogs at short to intermediate time scales where there are not too many overlapping aftershock sequences. For the spontaneous events, the interevent time is simply assumed to follow an exponential distribution with a constant rate, and asymmetry is not expected in this (Poisson process) case. In the following we show that the degree of asymmetry changes when considering the lagged interevent times.

%


\section{Materials and Methods}
 \subsection{Asymmetry}
Based on earthquake catalogs, we consider seismic events above a certain magnitude threshold (i.e. the magnitude of completeness for the given catalogue). For this sequence, we define the time interval between two successive earthquake events $i + 1$ and $i$ as the interevent time $\tau_i$ (in days). The lagged interevent time increment is defined as $\Delta \tau_i^{(k)}=\tau_{i+k}-\tau_{i}$ for a lag $k$ where $k$ is a positive integer lag. Following the above, the asymmetry measure of interevent times is defined as the ratio between the number of positive interevent time increments, $N_p$, minus the number of negative increments, $N_n$, and the total (positive plus negative) increments \cite{Ashkenazy2008}:
\begin{equation}\label{asy}
U(k)=\frac{N_p-N_n}{N_p+N_n}=\frac{\sum_i\Theta(\tau_{i+k}-\tau_{i})-\sum_i\Theta(\tau_{i}-\tau_{i+k})}{\sum_i\Theta(\tau_{i+k}-\tau_{i})+\sum_i\Theta(\tau_{i}-\tau_{i+k})},
\end{equation} 
where $\Theta(\tau)=1$ when $\tau>0$ and otherwise it is zero. We exclude the zero increments $\Delta\tau_i^{(k)}=0$ from the calculation; the number of zero increments is indeed very small. $U$ is bounded between -1 (monotonically decreasing sequence) and 1 (monotonically increasing sequence). When $U$ is close to zero, the time series is symmetric (for example, the PDF is nearly symmetrical close to $\Delta \tau^{(k)}=0$ for $k=1$ but highly asymmetric for $k=10$ and $k=50$ in Figure 2(a)). For instance, if the asymmetry value of $U=1/3$, the number of positive increments $N_p$ is twice the number of negative increments $N_n$ (i.e., $N_p$=2$N_n$). The positive (negative) increment of the interevent time represents the decreasing (increasing) earthquake rate. Similarly, we define the asymmetry of the interevent distance $r_i$ (in km) between the epicenters. 


\subsection{Generalized ETAS model}

We also study the asymmetry of synthetic catalogs based on the ETAS model in comparison to the asymmetry observed in the time series of real records. We use the ETAS model as a null hypothesis 
, since it is the most widely used statistical model to simulate the spatiotemporal clustering of seismic events \cite{Ogata1988, Ogata1998}. The earthquake sequence in the ETAS is defined as a stochastic Hawkes (point) process. We use the Gutenberg--Richter law (where $b=1$, truncated at $M_{max}$) to independently generate the magnitude of each earthquake ($\geq M_0$). For the ETAS model, the conditional intensity function $\lambda$ (which is basically the rate of earthquakes) at time $t$ with the seismic history $H_t$ prior to $t$ is given by 
\begin{equation}\label{lambda}
\lambda\left(t | H_t\right)= \mu + \sum_{i:t_i<t}f\left(m_i, t-t_i\right)\;,
\end{equation}
where $\mu$ is the background rate to generate spontaneous earthquakes estimated from the real catalogs \cite{Zhuang2010,Zhuang2012}.
The occurrence times of the past events are represented as $t_i$, and their magnitudes are $M_{i}$ ($\geq M_0$). Future earthquakes can be triggered by each past earthquake according to the generalized triggering function which here includes two triggering processes \cite{Zhang2020}, as   
\begin{equation}
f\left(M_i,t-t_i\right)=
\begin{cases}
\frac{Ac^p\exp[\alpha_1(m_i-M_0)]}{(t-t_i+c)^{p}} &  i > n-n_c\\
\frac{Ac^p\exp[\alpha_2(m_i-M_0)]}{(t-t_i+c)^{p}} &  i \leq n-n_c\;
\end{cases},
\label{Omori-Utsu}
\end{equation}
where $n-1$ is the total number of past events. The productivity of triggering earthquakes is controlled by the two productivity parameters $\alpha_{1}$ and $\alpha_{2}$ corresponding to the short-term ($i > n-n_c$) and long-term ($ i \leq n-n_c$) triggering respectively, which satisfy $\alpha_1\geq \alpha_2$. If the interevent number $n-i$ is smaller than the crossover number $n_c$, the $n$-th earthquake can be trigged by the $i$-th historical earthquake with a higher rate according to the larger $\alpha_1$. The crossover number $n_c$ is equal to $h10^{-bM_0}$ which is estimated from the memory measure of real earthquake catalog as reported by \citeA{Zhang2020}.  Also, we use the parameters $\{A, c, p, h, \alpha_{1}, \alpha_{2}\}$ in Eq. (\ref{Omori-Utsu}) estimated from real earthquake catalogs \cite{Zhang2020}. The generalized ETAS model reduces to the standard ETAS model if $\alpha_1=\alpha_2$. We add only two parameters, $\alpha_2$ and $n_c$ to the standard ETAS model and all other parameters remain the same. Note that when $\alpha_1$ is different with $\alpha_2$, Eq. \ref{Omori-Utsu} is a discontinuous function. Yet, it is very hard to observe a systematical discontinuity in synthetic catalogs of the generalized ETAS model as well as real data, since the cascading triggering process of aftershocks can weaken the discontinuity \cite{Zhang2020}.     
%
\subsection{Data} 
We analyze the Italian earthquake catalog between 1981 and 2017 \cite{Gasperini2013}. We also analyzed the Japan Unified High-Resolution Relocated Catalog for Earthquakes (JUICE) bewteen 2001 to 2012 \cite{Yano2017} and the Southern California catalog from 1981 to 2018 \cite{Hauksson2012}. The three catalogs are complete for magnitude threshold $3.0$ \cite{Hauksson2012,Gasperini2013,Yano2017} and this is also shown in Figure S1 where the distributions of the magnitudes ($\ge 3$) follow the Gutenberg-Richter law. 

\section{Results}   
\begin{figure}
\centering
\noindent\includegraphics[scale=0.7]{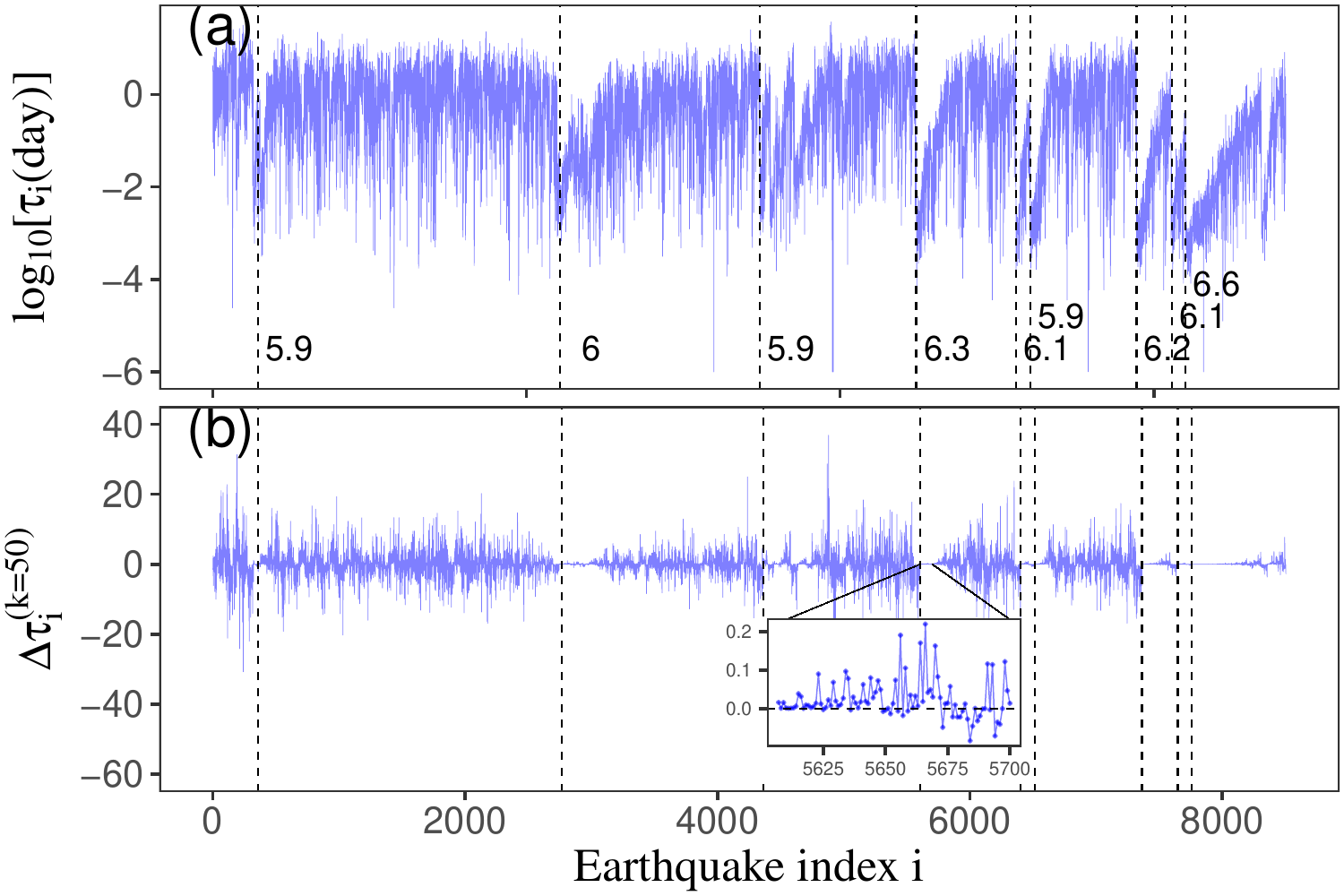}
\caption{Time series of (a) interevent times (log scale) and (b) their increments for $k$=50 for the Italian catalog (1981--2017) using magnitude threshold $3.0$. The inset figure shows the increments immediately after a large shock, indicating large asymmetry even for $k=50$. The black dashed vertical lines show the large earthquakes (magnitudes $\geq5.8$). Note the episodes of very small $\Delta \tau^{(k=50)}$ after large earthquakes; yet, these occur during very short time and will hardly visible when plotting $\Delta \tau^{(k=50)}$ versus time.}
\label{fig1}
\end{figure}

First, we obtain interevent times of the Italian earthquake catalog (using the threshold of magnitude $M_0=3.0$) and their increments for the lag index $k=50$. The results are shown in Fig. \ref{fig1}. As can be seen, interevent times decrease abruptly and then increase gradually after the occurrence of a large earthquake (Fig \ref{fig1}(a)), consistent with the Omori law. The increments are very small immediately after large shocks even for $k=50$, and most of them are positive (see inset figure) in Fig. \ref{fig1}(b). After sufficient time from a main shock, a crossover time, the rate of aftershocks decreases and the interevent increments become symmetric and switch between negative and positive values. 
Moreover, Fig. \ref{fig1}(b) shows that the interevent time increments for lag $k=50$ tend to be negative before the occurrence of large earthquakes. This observation can be explained as follows. The event time intervals before the main shock are relatively large in comparison to the event time intervals after the main shock, since the earthquake (aftershock) rate after the main shock is high. The difference between time interval for lag $k=50$, $\Delta\tau_i^{(k=50)}$, as approaching the main shock, involves the subtraction of a long-time interval before the main shock from a short time interval after the main shock, leading to negative $\Delta\tau_i^{(k=50)}$. Thus, $\Delta\tau_i^{(k=50)}$ will be negative $50$ lags preceding the main shock.

\begin{figure}
\centering
\noindent\includegraphics[scale=0.6]{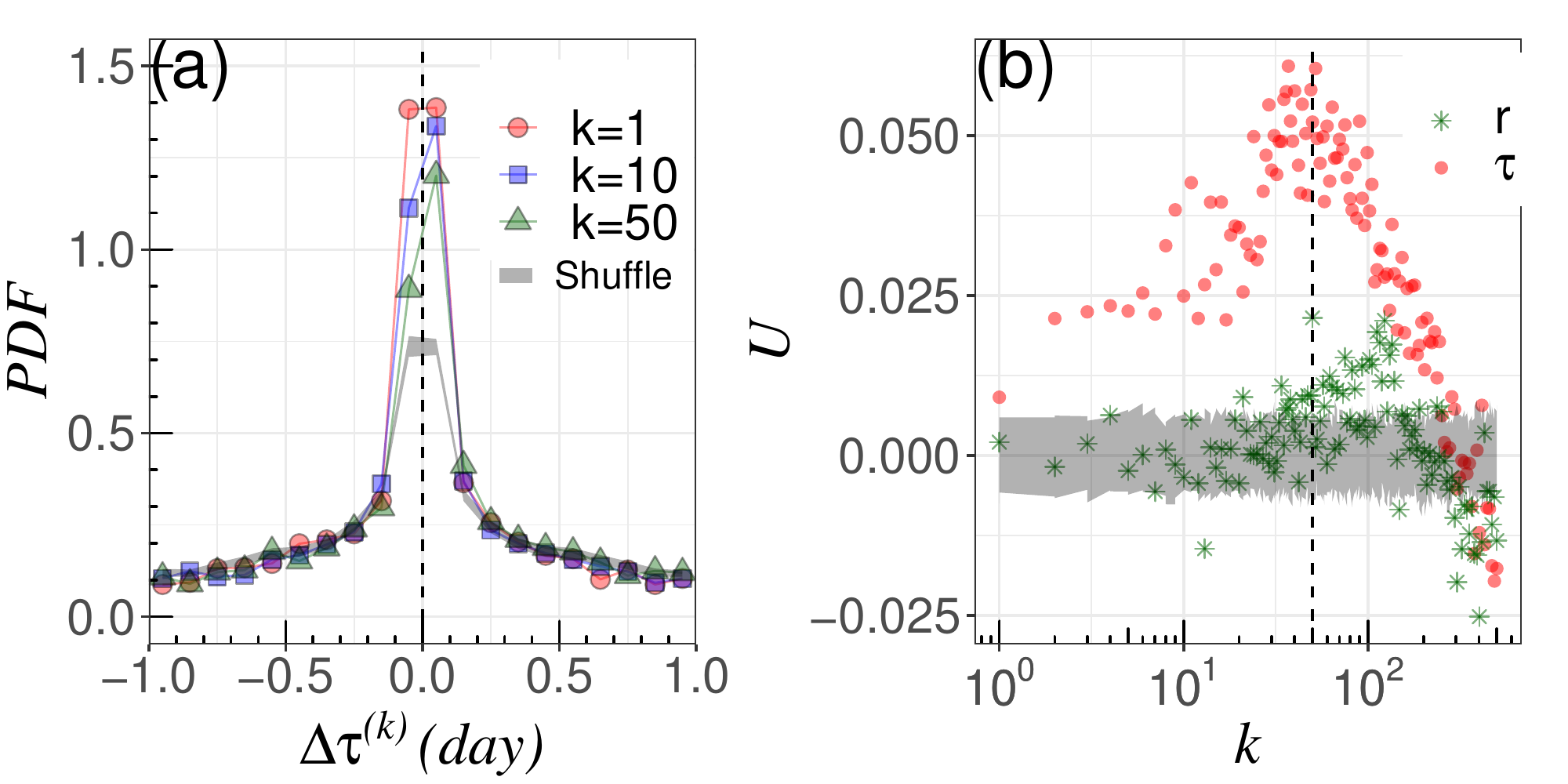}
\caption{(Color online) (a) PDF of interevent time increments $\Delta \tau^{(k)}$ for different lag index $k$ for the Italian catalog with the threshold  $M_0=3.0$. (b) The asymmetry measure $U$ versus the index $k$ for the interevent times (red) $\tau$ and distances (green) $r$ for the Italian catalog with $M_0=3.0$. Gray shades show the results of the shuffled (randomized) interevent times and distances and their standard deviations; these overlap each other. The dashed black line in (b) indicates the crossover lag $k_c\approx 50$ at which the asymmetry measure $U$ for interevent times is maximal.}
\label{fig2}
\end{figure}

Fig. \ref{fig2}(a) shows the Probability Density Function (PDF) of the interevent time increments for different $k$. The PDFs are essentially asymmetric about $\Delta \tau^{(k)}=0$ when $k$ increases and the asymmetry is dominated by the points close to $\Delta \tau^{(k)}=0$. To verify the significance of the asymmetry, we randomly shuffled the time series of interevent times and produced 100 shuffled sequences. This shuffling procedure destroys the (temporal) aftershock clustering such that the number of positive and negative increments should be similar. As a result, the number of small increments decreases and the number of large increments increases and the peak of the PDF (see gray shades in Fig. \ref{fig2}(a)) is much lower than the peak of the PDF of the original catalog. For a larger increment lag of $k=10$, $50$, the PDF becomes more asymmetric in comparison to PDF with $k=1$ and the PDF of the shuffled data [Fig.\ref{fig2}(a)]. More positive increments are observed for the larger lag-index $k$. 

To quantify the level of the asymmetry, we calculate the measure $U$ as a function of lag-index $k$ using Eq. (\ref{asy}). Fig. \ref{fig2}(b) depicts, for the Italian catalog, the measure $U$ for the interevent 
times (red) as a function of the lag increment $k$. The asymmetry measure, $U$, increases with $k$ for $k$ below a crossover lag, $k_c\approx 50$, and decreases with $k$ above the crossover $k_c$; $U$ is maximal at the crossover $k_c$ until it is indistinguishable from the random process shown at high $k$. As discussed above, we expect the presence of asymmetry in the interevent times following the Omori-law [Fig. \ref{fig1} and Fig. \ref{fig2}(a)]. Yet, the non-monotonic behavior with maximal asymmetry at $k_c$ is not trivial which implies a transition between the effect of the Omori-law and a random process. We also calculated the asymmetry measure, $U$, for interevent distances (green symbols in Fig. \ref{fig2}(b)) and observed similar behavior as for the interevent times, although much less pronounced. The results of shuffled, symmetric, time series (gray shaded area) are also included in Fig. \ref{fig2}(b) and indicate significant asymmetry for interevent times compared to this null hypothesis over a wide range of lags ($k<300$). For interevent distances we observe weak asymmetry only around lag $k=100$. 

\begin{figure}
\centering
\noindent\includegraphics[scale=0.5]{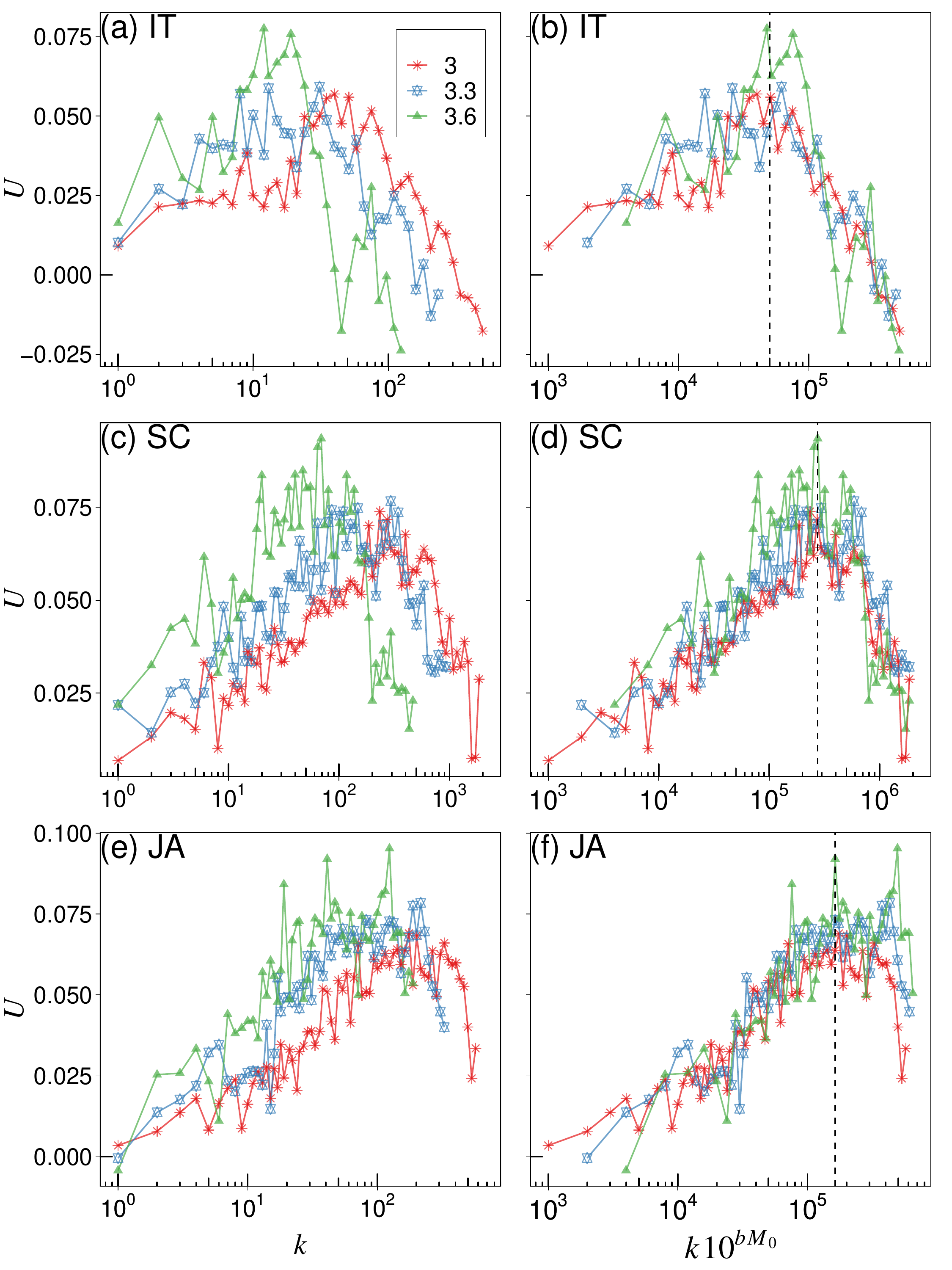}
\caption{(Color online) The asymmetry measure $U$ versus the index $k$ for the interevent times with different magnitude thresholds $M_0=3$, $3.3$ and $3.6$ for the earthquake catalogs of (a) Italy (IT),  (c) Southern California (SC) and (e) Japan (JA). (b), (d) and (f) Same as (a), (c) and (e) but $x$-axis is rescaled as $k10^{M_0}$. Dashed black lines indicate the approximate crossover lag.}
\label{fig3}
\end{figure}

We also calculated the asymmetry measure, $U$, for three magnitude thresholds and three places. Fig. \ref{fig3}(a), (c) and (e) show $U$ versus the index $k$ for the interevent times when using different magnitude thresholds $M_0=3$, $3.3$ and $3.6$ for the catalogs of Italy (IT), Southern California (SC) and Japan (JA). The asymmetry measure, $U$, exhibits similar increasing and decreasing trends for all three catalogs. The crossover lag, $k_c$, (at which the asymmetry is maximal) is smaller for the larger magnitude threshold (see Fig. \ref{fig3}(a), (c) and (e)). According to the Gutenberg-Richter law, the number of earthquakes decreases exponentially with the increasing magnitude threshold. Thus, we rescale the lag $k$ with $k10^{bM_0}$ and the results are shown in Fig. \ref{fig3}(b), (d) and (f). The different asymmetry curves collapse into a single curve for which the crossover is $k_c10^{bM_0}$; this scaling approach is similar to the scaling procedure of interevent times discussed in previous studies \cite{Bak2002,Corral2003,Saichev2006,Sornette2008}. However, the crossovers are not the same for different places. For IT, the rescaled crossover lag is, $k_c10^{bM_0}\approx 5\times 10^4$ and is smaller than the crossovers for JA ($k_c10^{bM_0}\approx 2\times 10^5$) and SC ($k_c10^{bM_0}\approx 3\times 10^5$). The asymmetry curves also satisfy the scaling relation by rescaling the lag $k$ with the averaged time intervals $\langle \tau \rangle$ as shown in Fig. S2. We thus obtain that the crossover times approximately correspond to 80, 280, and 50 days for IT, SC and JA respectively. We also consider and observed the asymmetry for different region sizes as shown in Fig. S3(a). A smaller region size shows a larger asymmetry since more events (aftershocks) are correlated within the area as proposed by \citeA{Touati2009}. Moreover, the crossover can be scaled with respect to region size in Fig. S3(b). Figure S4 shows the weak asymmetry for the global earthquake catalog.

\begin{figure}
\centering
\noindent\includegraphics[scale=0.15]{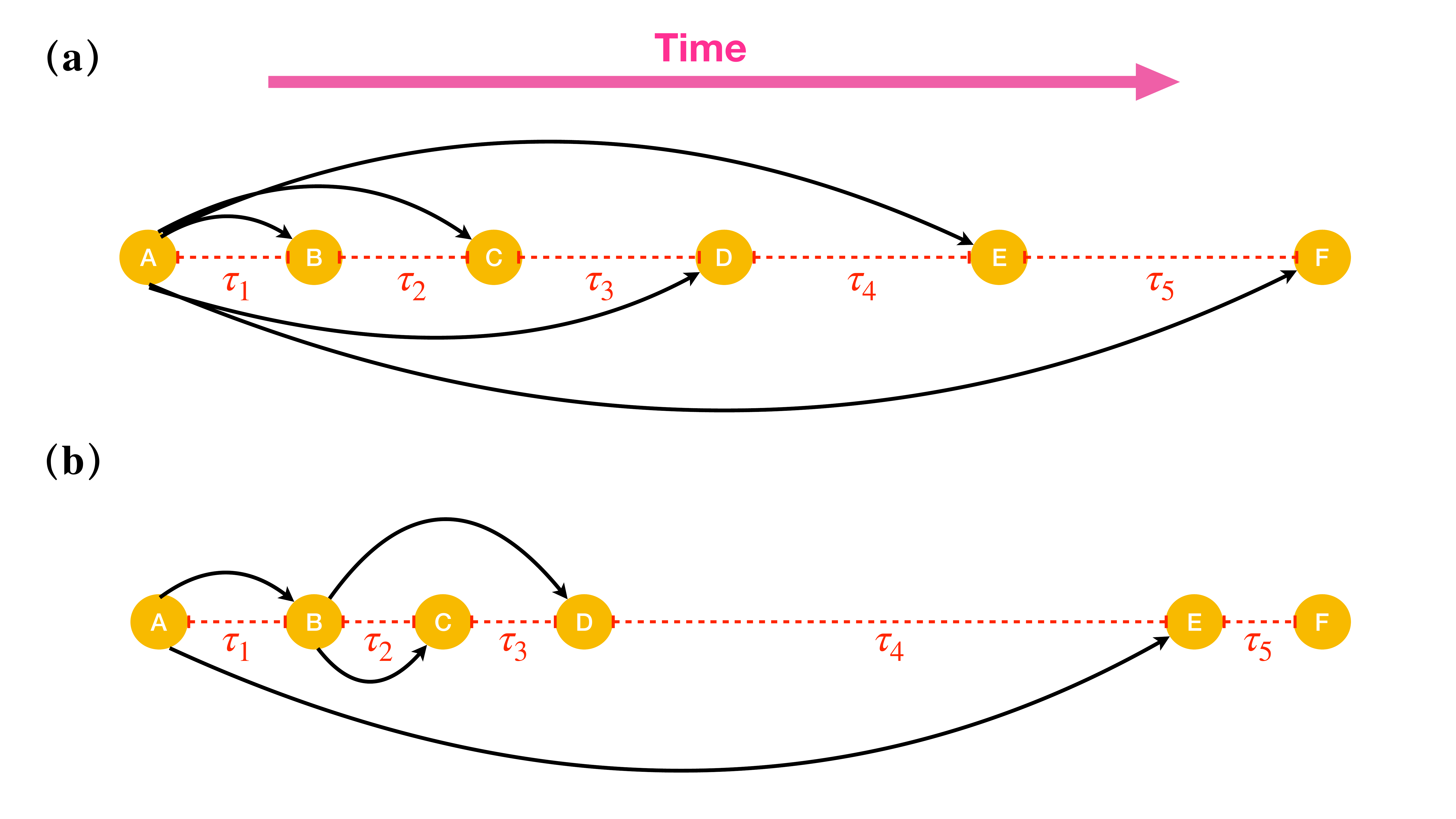}
\caption{Cartoon illustrating how interevent times in earthquakes change with time for (a) the aftershocks $B$--$F$ directly triggered by $A$ (mainshock), and (b) the first generation of aftershocks $B$, and $E$ triggered by $A$, the second generation of aftershocks $C$, and $D$ triggered by $B$ and the spontaneous event $F$. Dashed red lines represent interevent times.}
\label{fig4}
\end{figure}

We now aim to explain the mechanism underlying the observed asymmetry measure. Considering a simple situation, for which aftershocks $B$--$F$ are the first generation aftershocks triggered by a mainshock $A$, as shown by the schematic drawing in Fig. \ref{fig4}(a). Due to the Omori law, the frequency of aftershocks decreases like $(t-t_0)^{-p}$ ($p$ is close to 1 and $t-t_0$ is the time since the mainshock) such that the interevent time $\tau$ after the mainshock follows $\tau\sim (t-t_0)^p$. Thus, the interevent time statistically increases with time, resulting in a positive asymmetry with more positive increments in comparison to negative increments. However, the real situation is more complex as not only mainshocks can trigger aftershocks but aftershocks can also trigger other aftershocks. Moreover, spontaneous earthquakes (mainshocks) could be mixed with aftershocks due to the stacking involved (see the example in Fig. \ref{fig4}(b)). The indirect triggered events and the spontaneous events can decrease the interevent times as shown in Fig. \ref{fig4}(b) ($\tau_2$, $\tau_3$ and $\tau_5$ are smaller than $\tau_1$). The above considerations implies that the events below the crossover lag $k_c$ are mainly triggered by a mainshock. Above the crossover ($k>k_c$), the sequences for spontaneous and triggered events will overlap with high probabilities resulting in a fast decay of asymmetry. 


The ETAS model is widely used to simulate and study the temporal clustering of seismic events \cite{Ogata1988, Ogata1998}. The rate function of the ETAS model consists of the spontaneous (background) rate and triggering rate of historical events (see Eq. (\ref{lambda})). The choice of parameters in the ETAS model is critical to reproduce the features of real earthquake sequences. The maximum likelihood estimation (MLE) procedure has been proposed \cite{Zhuang2010} to estimate the parameters. In a recent study \cite{Zhang2020} the conventional ETAS model has been found to be unable to reproduce important (long-term) memory characteristics observed in real catalogs. In the same study a generalized the ETAS model has been developed and found to be useful in reproducing the observe memory features that appear in the real catalogs (see Materials and Methods). Below we test the asymmetry of the generalized ETAS model for Italy with three choices of parameters: (I) $\alpha_1=\alpha_2=\alpha$ such that the generalized model reduces to the standard ETAS model. The parameters are estimated using the MLE. This choice is termed ``EM0''. (II) Since some studies have reported that the $\alpha$-value is underestimated by the MLE \cite{Marzocchi2009,Seif2017,Zhuang2019}, we consider a second choice of parameters termed ``EM1'', which is the same as EM0 but with larger $\alpha$ (and smaller $A$ to guarantee the similar branching ratio) (Eq. (\ref{Omori-Utsu})). (III) The generalized ETAS model with $\alpha_1>\alpha_2$ developed recently \cite{Zhang2020}. This choice is termed ``EM2''. The selected parameters of EM0, EM1 and EM2 for the Italian catalog are listed in Table \ref{T1}. We generated 50 realizations of synthetic catalogs with magnitudes greater than or equal to magnitude 3, each covering 50000 days. The earthquake rates are 0.69$\pm$0.03, 0.73$\pm$0.1 and 0.71$\pm$0.06 events per day for EM0,  EM1 and EM2 respectively. The rates of the models are similar and close to that of the real data. 

\begin{table}
\caption{\label{T1}%
Estimated parameters of the three versions of the ETAS model for the Italian catalog. The parameters of EM0 are taken from \cite{Lombardi2015} which have been estimated by MLE. For EM1, the parameters $\alpha_1$, $\alpha_2$ are larger than EM0, and $A$ is smaller, to guarantee similar earthquake rate as the real catalog. We select the parameters of EM2 based on recent findings \cite{Zhang2020}.}
\centering
\begin{tabular}{cccccccc}
\toprule
\textrm{}&
\textrm{$\mu$}&
\textrm{$c$}&
\textrm{$p$}&
\textrm{$A$}&
\textrm{$\alpha_1$}&
\textrm{$\alpha_2$}&
\textrm{$h$}\\
\midrule
 EM0& 0.2 & 0.007 & 1.13 & 6.26 & 1.4 & 1.4 & --\\
 \midrule
 EM1& 0.2 & 0.007 & 1.13 & 2.91 & 2.0 & 2.0 & --\\
  \midrule
 EM2& 0.2 & 0.007 & 1.13 & 3.35 & 2.0 & 1.4 & $2\times10^5$\\
\bottomrule
\end{tabular}
\end{table}

\begin{figure}
\centering
\noindent\includegraphics[scale=0.4]{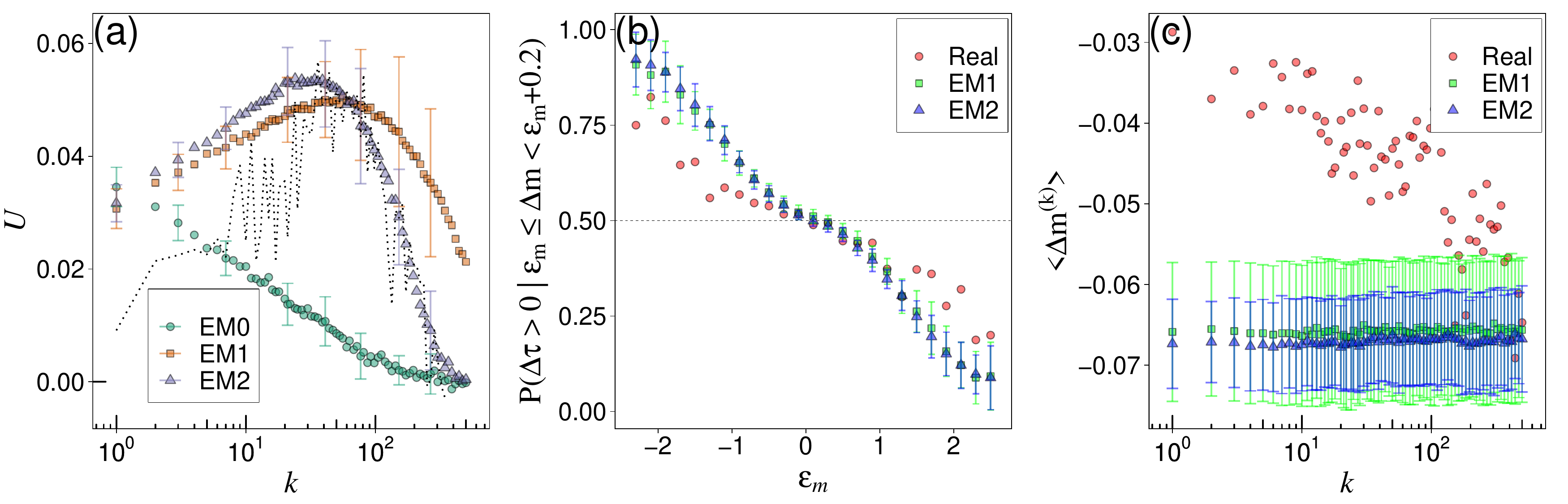}
\caption{(color online) (a) The asymmetry measure $U$ versus the index $k$ for the interevent times of the synthetic Italian catalogs using EM0, EM1 and EM2 with magnitude threshold 3. The asymmetry of the real Italian catalog is indicated by the dotted line. (b) Conditional probability $P(\Delta\tau>0|\epsilon_m\leq \Delta m<\epsilon_m+0.2)$ as a function of the magnitude increment $\epsilon_m$ for the real Italian catalog and synthetic catalogs of EM1, EM2. (c) The average of the magnitude increment $\langle\Delta m^{(k)}\rangle$ for $\tau_i<0.5$ days versus the index $k$. The measures and their error bars are calculated based on the means and the standard deviations of 50 independent realizations for the models.}
\label{fig5}
\end{figure}

Next, we study the asymmetry for the three versions of the ETAS model introduced above. Figure \ref{fig5}(a) shows that the asymmetry of the interevent times in the standard ETAS, EM0, in marked contrast with real asymmetry, deceases with the lag index $k$ without a crossover for EM0 (green dots). Both, EM1 (red squares) and EM2 (green triangles) exhibit much better performance and their asymmetry curves are similar to the real catalog (dotted line). Due to the smaller $\alpha$ in EM0 relative to EM1, the probability of aftershocks directly triggered by a large mainshock is too low to increase the asymmetry for EM0. Thus, the asymmetry deceases as the lag index $k$ increases at the beginning rather than after a certain lag. The asymmetry of EM0 demonstrates that the $\alpha$-value is indeed underestimated by MLE. Comparing between EM1 and EM2, the asymmetry of EM2 decays faster above the crossover, more similar to the decay of the real catalog, the dotted line (see Fig. \ref{fig5}(a)). Moreover, the crossover point is different for EM1 ($k_c\approx 60$) and EM2 ($k_c\approx 50$). Thus, the crossover of EM2 is closer to the observed one (Fig. \ref{fig5}(a)). We thus conclude that the two-alpha ($\alpha_1>\alpha_2$) ETAS model exhibits the best performance in reproducing both the memory \cite{Zhang2020} and asymmetry in the current study than both versions of the standard ETAS model. The asymmetry of EM2 also satisfies the scaling relation for the magnitude threshold similar to the real one (see SI, Figure S5).

We further study the dependence of the interevent time increments $\Delta \tau^k_i$ on the magnitude increment $\Delta m_i^{(k)}=m_{i+k}-m_i$, to understand in more details the role of Omori law on the asymmetry. For this purpose we calculated the conditional probability $P(\Delta \tau>0|\epsilon_m\leq \Delta m<\epsilon_m+d)$, where $d=0.2$ is the bin size of the magnitude increment. Figure \ref{fig5}(b) shows that this conditional probability decreases when the magnitude increment $\epsilon_m$ increases, for the real, EM1 and EM2 catalogs;. Moreover, the conditional probability is around 0.5 (corresponding to the asymmetry measure around zero) when $\epsilon_m$ is close to zero. Thus, the asymmetry measure close to zero could be due to the magnitude similarity for small lag-index $k$, as the PDF interevent time increments is maximal close to zero (Fig \ref{fig2}(a)). Previous studies \cite{Lippiello2008,Lippiello2012} have found that the magnitude of consecutive events is more similar than would be expected from random sampling of the Gutenber-Richter distribution. It is apparent that EM1 and EM2 overestimate the conditional probability of the real data. Fig. \ref{fig5}(c) shows the average of the magnitude increment $\langle\Delta m^{(k)}\rangle$ for $\tau_i<0.5$ days (to focus on aftershocks) as a function of the lag-index $k$. The size of aftershock is usually smaller than the mainshock yielding the negative values in Figure \ref{fig5}(c). Yet, it is also clear that while the mean magnitude difference $\langle\Delta m^{(k)}\rangle$ is almost constant with lag $k$ for EM1 and EM2, it decreases for the real catalog, from values closer to zero for small lag-index $k$ to values of EM1 and EM2 at large $k$ (Fig. \ref{fig5}(c)). These results indicate that the magnitude similarity reported by \cite{Lippiello2008,Lippiello2012} is absent in both models.

While the asymmetry of EM2 is similar to the asymmetry of the real catalog for $k>300$, it is significantly higher for smaller $k$ (Fig. \ref{fig5}(a)). The Italian catalog we used has been reported to be complete above magnitude threshold 3.0. \cite{Gasperini2013}. However, due to the inefficiency of the seismic network and the overlapping of aftershock seismograms, an earthquake catalog could be incomplete, especially after mainshocks \cite{Kagan2004,Hainzl2016,DeArcangelis2018}. To investigate the effect of the incompleteness of the catalogs, we generate synthetic incomplete catalogs based on the studies of \citeA{helmstetter2006comparison,Seif2017,petrillo2021testing}. The incomplete ETAS model is based on the conditional earthquake rate intensity function as \cite{petrillo2021testing}, 
\begin{equation}\label{inetas}
 \lambda_I\left(m,t\right)=\lambda\times\prod_{i}{\Phi\left(m\middle|M_i\left(t-t_i\right),\sigma\right)},
\end{equation} 
where all past events with $t_i < t$ are considered. We define $\Phi\left(m\middle|M_i\left(t-t_i\right),\sigma\right)=1$ when $m>M_i\left(t-t_i\right)+\sigma$, $\Phi\left(m\middle|M_i\left(t-t_i\right),\sigma\right)=0$ when $m<M_i\left(t-t_j\right)-\sigma$,  $\Phi\left(m\middle|M_i\left(t-t_i\right),\sigma\right)=0.5$ else. The magnitude threshold $M_i\left(t-t_i\right)$ is calculated as \cite{Kagan2004,Hainzl2016,DeArcangelis2018},
 \begin{equation}\label{inmgth}
M_i\left(t-t_i\right)=m_i-\delta_0-\omega log_{10}\left(t-t_i\right),
\end{equation}  
where $m_i$ is the magnitude of past event $i$, and $t-t_i$ is the time since the past event. The parameter $\sigma=0.6$ is chosen following \citeA{petrillo2021testing} and \citeA{Seif2017,helmstetter2006comparison} suggested the following parameter values $\delta_0=4.5$ and $\omega=0.75$. We consider three different choices of the parameter $\delta_0$, $\delta_0=4.5$, $4.0$, and $3.5$ to generate synthetic catalogs with different degree of incompleteness.

\begin{figure}
\centering
\noindent\includegraphics[scale=0.4]{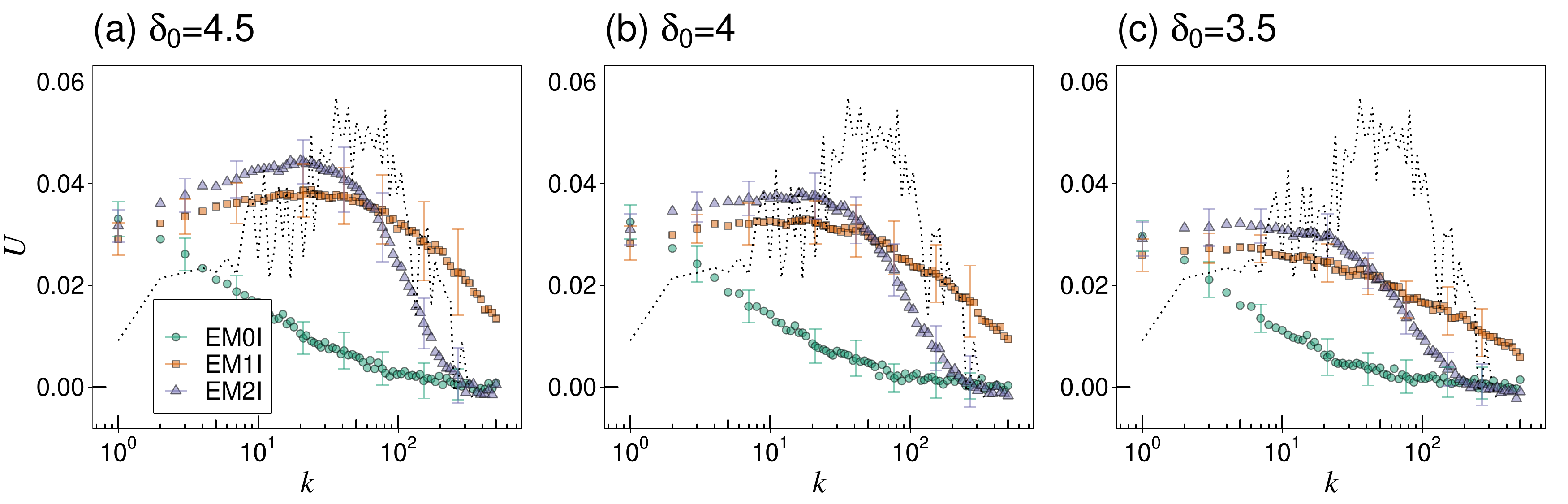}
\caption{(Color online) The asymmetry measure $U$ versus the index $k$ for the interevent times of the synthetic incomplete catalogs using EM0I, EM1I and EM2I with magnitude threshold 3 for (a) $\delta_0=4.5$ (EM0I with $\alpha_1=\alpha_2=1.41$, EM1I with $\alpha_1=\alpha_2=2.01$, and EM2I with $\alpha_1=2.01$ and $\alpha_2=1.40$), (b) $\delta_0=4.0$ (EM0I with $\alpha_1=\alpha_2=1.42$, EM1I with $\alpha_1=\alpha_2=2.02$, and EM2I with $\alpha_1=2.02$ and $\alpha_2=1.40$) and (c) $\delta_0=3.5$ (EM0I with $\alpha_1=\alpha_2=1.44$, EM1I with $\alpha_1=\alpha_2=2.04$, and EM2I with $\alpha_1=2.04$ and $\alpha_2=1.40$). The asymmetry of the real Italian catalog is indicated by the dotted line.}
\label{fig6}
\end{figure}

To generate the synthetic incomplete catalogs based on EM0, EM1 and EM2 (represented as EM0I, EM1I and EM2I respectively), we remove an aftershock $i$ from the synthetic complete catalogs with a probability given by $\prod_{i}{\Phi\left(m\middle|M_i\left(t-t_i\right),\sigma\right)}$ \cite{petrillo2021testing}. To roughly preserve the total number of earthquakes to be the same as that of Figure \ref{fig5}(a), we increased slightly the $\alpha$ parameter and left the other parameters unchanged. With this procedure, the level of incompleteness of each synthetic catalog was 5\%, 10\% and 20\% for $\delta_0=4.5$ , $4.0$, and $3.5$ respectively for EM1I and EM2I; the percentages indicate the relative number of events that has been removed from the complete catalog. It is apparent from our results (see Figure \ref{fig6}) that the asymmetry weakens as the degree of incompleteness is higher. Still, for both models, the asymmetry is overestimated for small lag index $k$ in comparison to the real catalog and weakens when the catalogs are more incomplete. Figure S6 shows similar results when using $\sigma=0.6$, $1.2$, $1.8$ to control the degree of incompleteness. We also try to control the parameter $A$ to keep the same number of earthquakes for EM0I, EM1I and EM2I and the results are shown in Figure S7.

\section{Conclusions}
Here, we investigated the asymmetry behavior of interevent times (and distances) in earthquake catalogs. For real seismic catalogs, the asymmetry as a function of $k$ first increases up to a crossover lag $k_c$ and then decreases rapidly. The crossover lag $k_c$ changes with location and with the magnitude threshold, where the latter can be rescaled to unified value of $k_c10^{bM_0}$. We suggest that the Omori law is associated with the increase of the asymmetry below the crossover and has a decreasing influence above this crossover. This is probably due to the overlapping of different triggered aftershocks and the spontaneous events that lead to a fast decay of asymmetry above the crossover. The de-clustering between spontaneous and triggered earthquake events is still an open important question \cite{Zaliapin2008,Yehuda2013}. The asymmetry results reported here and its associated crossover may help to resolve this question although this requires further investigation.                  

In the standard ETAS model whose parameters are estimated by MLE, the increase of asymmetry and the crossover cannot be reproduced. When the $\alpha$-value is increased, a large mainshock can trigger more aftershocks such that there exists an increasing trend and a crossover in the standard ETAS model. This demonstrates that the common $\alpha$-value is indeed underestimated by MLE. However, the crossover value of $k$ is larger and the asymmetry above the crossover is significantly higher and decays slower in the standard ETAS model with large $\alpha$ than the real one. The generalized ETAS model with two $\alpha$-values ($\alpha_1>\alpha_2$) in short and long time scales exhibits similar asymmetry behavior as that of the real catalog for lags larger than the crossover lag $k_c$. Yet, the asymmetry for small lag-index $k$ is overestimated by both models (one and two $\alpha$-value). We suggest that the short-term symmetrical behavior can be attributed to the magnitude similarity in real data which is missing in both models. The additional advantage of the generalized ETAS model is its ability to reproduce the observe memory in earthquake catalogs as reported in \cite{Zhang2020}. Thus, generally speaking, the asymmetry findings reported here may be used to improve earthquake forecasting models as the asymmetry measure can serve as an additional characteristic that a forecasting model should reproduce.

\acknowledgments
We thank for the financial support by the EU H2020 project RISE, the Israel Science Foundation (Grants No. 189/19), DTRA ,the Pazy Foundation, the joint China-Israel Science Foundation (Grants No. 3132/19) and the BIU Center for Research in Applied Cryptography and Cyber Security. We thank the Israel ministry of energy. We downloaded the Southern California catalog from the SCEDC (https://scedc.caltech.edu/research-tools/alt-2011-dd-hauksson-yang-shearer.html) \cite{Hauksson2012} and the Japanese Catalog (JUICE) from ref \cite{Yano2017}. The Italian catalog is available on request from ref \cite{Gasperini2013} and the authors. 
 
%
%

\bibliography{earthquake}

\begin{thebibliography}{}

\bibitem [\protect \citeauthoryear {%
An%
}{%
An%
}{%
{\protect \APACyear {2004}}%
}]{%
An2004}
\APACinsertmetastar {%
An2004}%
\begin{APACrefauthors}%
An, S\BPBI I.%
\end{APACrefauthors}%
\unskip\
\newblock
\APACrefYearMonthDay{2004}{}{}.
\newblock
{\BBOQ}\APACrefatitle {{Interdecadal changes in the El Nino-La Nina asymmetry}}
  {{Interdecadal changes in the El Nino-La Nina asymmetry}}.{\BBCQ}
\newblock
\APACjournalVolNumPages{Geophys. Res. Lett.}{31}{23}{1--4}.
\newblock
\begin{APACrefDOI} \doi{10.1029/2004GL021699} \end{APACrefDOI}
\PrintBackRefs{\CurrentBib}

\bibitem [\protect \citeauthoryear {%
Ashkenazy%
, Feliks%
, Gildor%
\BCBL {}\ \BBA {} Tziperman%
}{%
Ashkenazy%
\ \protect \BOthers {.}}{%
{\protect \APACyear {2008}}%
}]{%
Ashkenazy2008}
\APACinsertmetastar {%
Ashkenazy2008}%
\begin{APACrefauthors}%
Ashkenazy, Y.%
, Feliks, Y.%
, Gildor, H.%
\BCBL {}\ \BBA {} Tziperman, E.%
\end{APACrefauthors}%
\unskip\
\newblock
\APACrefYearMonthDay{2008}{}{}.
\newblock
{\BBOQ}\APACrefatitle {{Asymmetry of daily temperature records}} {{Asymmetry of
  daily temperature records}}.{\BBCQ}
\newblock
\APACjournalVolNumPages{J. Atmos. Sci.}{65}{10}{3327--3336}.
\newblock
\begin{APACrefDOI} \doi{10.1175/2008JAS2662.1} \end{APACrefDOI}
\PrintBackRefs{\CurrentBib}

\bibitem [\protect \citeauthoryear {%
Bak%
, Christensen%
, Danon%
\BCBL {}\ \BBA {} Scanlon%
}{%
Bak%
\ \protect \BOthers {.}}{%
{\protect \APACyear {2002}}%
}]{%
Bak2002}
\APACinsertmetastar {%
Bak2002}%
\begin{APACrefauthors}%
Bak, P.%
, Christensen, K.%
, Danon, L.%
\BCBL {}\ \BBA {} Scanlon, T.%
\end{APACrefauthors}%
\unskip\
\newblock
\APACrefYearMonthDay{2002}{}{}.
\newblock
{\BBOQ}\APACrefatitle {{Unified Scaling Law for Earthquakes}} {{Unified Scaling
  Law for Earthquakes}}.{\BBCQ}
\newblock
\APACjournalVolNumPages{Phys. Rev. Lett.}{88}{17}{178501}.
\newblock
\begin{APACrefDOI} \doi{10.1103/PhysRevLett.88.178501} \end{APACrefDOI}
\PrintBackRefs{\CurrentBib}

\bibitem [\protect \citeauthoryear {%
Corral%
}{%
Corral%
}{%
{\protect \APACyear {2003}}%
}]{%
Corral2003}
\APACinsertmetastar {%
Corral2003}%
\begin{APACrefauthors}%
Corral, {\'{A}}.%
\end{APACrefauthors}%
\unskip\
\newblock
\APACrefYearMonthDay{2003}{}{}.
\newblock
{\BBOQ}\APACrefatitle {{Local distributions and rate fluctuations in a unified
  scaling law for earthquakes}} {{Local distributions and rate fluctuations in
  a unified scaling law for earthquakes}}.{\BBCQ}
\newblock
\APACjournalVolNumPages{Phys. Rev. E}{68}{3}{035102}.
\newblock
\begin{APACrefDOI} \doi{10.1103/PhysRevE.68.035102} \end{APACrefDOI}
\PrintBackRefs{\CurrentBib}

\bibitem [\protect \citeauthoryear {%
Corral%
}{%
Corral%
}{%
{\protect \APACyear {2004}}%
}]{%
Corral2003a}
\APACinsertmetastar {%
Corral2003a}%
\begin{APACrefauthors}%
Corral, {\'{A}}.%
\end{APACrefauthors}%
\unskip\
\newblock
\APACrefYearMonthDay{2004}{}{}.
\newblock
{\BBOQ}\APACrefatitle {{Long-Term Clustering, Scaling, and Universality in the
  Temporal Occurrence of Earthquakes}} {{Long-Term Clustering, Scaling, and
  Universality in the Temporal Occurrence of Earthquakes}}.{\BBCQ}
\newblock
\APACjournalVolNumPages{Phys. Rev. Lett.}{92}{10}{108501}.
\newblock
\begin{APACrefDOI} \doi{10.1103/PhysRevLett.92.108501} \end{APACrefDOI}
\PrintBackRefs{\CurrentBib}

\bibitem [\protect \citeauthoryear {%
Davidsen%
, Stanchits%
\BCBL {}\ \BBA {} Dresen%
}{%
Davidsen%
\ \protect \BOthers {.}}{%
{\protect \APACyear {2007}}%
}]{%
Davidsen2007}
\APACinsertmetastar {%
Davidsen2007}%
\begin{APACrefauthors}%
Davidsen, J.%
, Stanchits, S.%
\BCBL {}\ \BBA {} Dresen, G.%
\end{APACrefauthors}%
\unskip\
\newblock
\APACrefYearMonthDay{2007}{}{}.
\newblock
{\BBOQ}\APACrefatitle {{Scaling and universality in rock fracture}} {{Scaling
  and universality in rock fracture}}.{\BBCQ}
\newblock
\APACjournalVolNumPages{Phys. Rev. Lett.}{98}{12}{125502}.
\newblock
\begin{APACrefDOI} \doi{10.1103/PhysRevLett.98.125502} \end{APACrefDOI}
\PrintBackRefs{\CurrentBib}

\bibitem [\protect \citeauthoryear {%
de Arcangelis%
, Godano%
, Grasso%
\BCBL {}\ \BBA {} Lippiello%
}{%
de Arcangelis%
\ \protect \BOthers {.}}{%
{\protect \APACyear {2016}}%
}]{%
DeArcangelis2016}
\APACinsertmetastar {%
DeArcangelis2016}%
\begin{APACrefauthors}%
de Arcangelis, L.%
, Godano, C.%
, Grasso, J\BPBI R.%
\BCBL {}\ \BBA {} Lippiello, E.%
\end{APACrefauthors}%
\unskip\
\newblock
\APACrefYearMonthDay{2016}{}{}.
\newblock
{\BBOQ}\APACrefatitle {{Statistical physics approach to earthquake occurrence
  and forecasting}} {{Statistical physics approach to earthquake occurrence and
  forecasting}}.{\BBCQ}
\newblock
\APACjournalVolNumPages{Phys. Rep.}{628}{}{1--91}.
\newblock
\begin{APACrefDOI} \doi{10.1016/j.physrep.2016.03.002} \end{APACrefDOI}
\PrintBackRefs{\CurrentBib}

\bibitem [\protect \citeauthoryear {%
de Arcangelis%
, Godano%
\BCBL {}\ \BBA {} Lippiello%
}{%
de Arcangelis%
\ \protect \BOthers {.}}{%
{\protect \APACyear {2018}}%
}]{%
DeArcangelis2018}
\APACinsertmetastar {%
DeArcangelis2018}%
\begin{APACrefauthors}%
de Arcangelis, L.%
, Godano, C.%
\BCBL {}\ \BBA {} Lippiello, E.%
\end{APACrefauthors}%
\unskip\
\newblock
\APACrefYearMonthDay{2018}{}{}.
\newblock
{\BBOQ}\APACrefatitle {{The Overlap of Aftershock Coda Waves and Short-Term
  Postseismic Forecasting}} {{The Overlap of Aftershock Coda Waves and
  Short-Term Postseismic Forecasting}}.{\BBCQ}
\newblock
\APACjournalVolNumPages{J. Geophys. Res. Solid Earth}{123}{7}{5661--5674}.
\newblock
\begin{APACrefDOI} \doi{10.1029/2018JB015518} \end{APACrefDOI}
\PrintBackRefs{\CurrentBib}

\bibitem [\protect \citeauthoryear {%
Fan%
\ \protect \BOthers {.}}{%
Fan%
\ \protect \BOthers {.}}{%
{\protect \APACyear {2019}}%
}]{%
Fan2018b}
\APACinsertmetastar {%
Fan2018b}%
\begin{APACrefauthors}%
Fan, J.%
, Zhou, D.%
, Shekhtman, L\BPBI M.%
, Shapira, A.%
, Hofstetter, R.%
, Marzocchi, W.%
\BDBL {}Havlin, S.%
\end{APACrefauthors}%
\unskip\
\newblock
\APACrefYearMonthDay{2019}{}{}.
\newblock
{\BBOQ}\APACrefatitle {{Possible origin of memory in earthquakes: Real catalogs
  and an epidemic-type aftershock sequence model}} {{Possible origin of memory
  in earthquakes: Real catalogs and an epidemic-type aftershock sequence
  model}}.{\BBCQ}
\newblock
\APACjournalVolNumPages{Phys. Rev. E}{99}{4}{042210}.
\newblock
\begin{APACrefDOI} \doi{10.1103/PhysRevE.99.042210} \end{APACrefDOI}
\PrintBackRefs{\CurrentBib}

\bibitem [\protect \citeauthoryear {%
Gasperini%
, Lolli%
\BCBL {}\ \BBA {} Vannucci%
}{%
Gasperini%
\ \protect \BOthers {.}}{%
{\protect \APACyear {2013}}%
}]{%
Gasperini2013}
\APACinsertmetastar {%
Gasperini2013}%
\begin{APACrefauthors}%
Gasperini, P.%
, Lolli, B.%
\BCBL {}\ \BBA {} Vannucci, G.%
\end{APACrefauthors}%
\unskip\
\newblock
\APACrefYearMonthDay{2013}{}{}.
\newblock
{\BBOQ}\APACrefatitle {{Empirical calibration of local magnitude data sets
  versus moment magnitude in Italy}} {{Empirical calibration of local magnitude
  data sets versus moment magnitude in Italy}}.{\BBCQ}
\newblock
\APACjournalVolNumPages{Bull. Seismol. Soc. Am.}{103}{4}{2227--2246}.
\newblock
\begin{APACrefDOI} \doi{10.1785/0120120356} \end{APACrefDOI}
\PrintBackRefs{\CurrentBib}

\bibitem [\protect \citeauthoryear {%
Gutenberg%
\ \BBA {} Richter%
}{%
Gutenberg%
\ \BBA {} Richter%
}{%
{\protect \APACyear {1944}}%
}]{%
Gutenberg1944a}
\APACinsertmetastar {%
Gutenberg1944a}%
\begin{APACrefauthors}%
Gutenberg, B.%
\BCBT {}\ \BBA {} Richter, C\BPBI F.%
\end{APACrefauthors}%
\unskip\
\newblock
\APACrefYearMonthDay{1944}{}{}.
\newblock
{\BBOQ}\APACrefatitle {{Frequency of Earthquakes in California}} {{Frequency of
  Earthquakes in California}}.{\BBCQ}
\newblock
\APACjournalVolNumPages{Bull. Seismol. Soc. Am.}{34}{4}{185--188}.
\newblock
\begin{APACrefDOI} \doi{10.1038/156371a0} \end{APACrefDOI}
\PrintBackRefs{\CurrentBib}

\bibitem [\protect \citeauthoryear {%
Hainzl%
}{%
Hainzl%
}{%
{\protect \APACyear {2016}}%
}]{%
Hainzl2016}
\APACinsertmetastar {%
Hainzl2016}%
\begin{APACrefauthors}%
Hainzl, S.%
\end{APACrefauthors}%
\unskip\
\newblock
\APACrefYearMonthDay{2016}{}{}.
\newblock
{\BBOQ}\APACrefatitle {{Rate‐Dependent Incompleteness of Earthquake
  Catalogs}} {{Rate‐Dependent Incompleteness of Earthquake Catalogs}}.{\BBCQ}
\newblock
\APACjournalVolNumPages{Seismol. Res. Lett.}{87}{2A}{337--344}.
\newblock
\begin{APACrefDOI} \doi{10.1785/0220150211} \end{APACrefDOI}
\PrintBackRefs{\CurrentBib}

\bibitem [\protect \citeauthoryear {%
Hauksson%
, Yang%
\BCBL {}\ \BBA {} Shearer%
}{%
Hauksson%
\ \protect \BOthers {.}}{%
{\protect \APACyear {2012}}%
}]{%
Hauksson2012}
\APACinsertmetastar {%
Hauksson2012}%
\begin{APACrefauthors}%
Hauksson, E.%
, Yang, W.%
\BCBL {}\ \BBA {} Shearer, P\BPBI M.%
\end{APACrefauthors}%
\unskip\
\newblock
\APACrefYearMonthDay{2012}{}{}.
\newblock
{\BBOQ}\APACrefatitle {{Waveform Relocated Earthquake Catalog for Southern
  California (1981 to June 2011)}} {{Waveform Relocated Earthquake Catalog for
  Southern California (1981 to June 2011)}}.{\BBCQ}
\newblock
\APACjournalVolNumPages{Bull. Seismol. Soc. Am.}{102}{5}{2239--2244}.
\newblock
\begin{APACrefDOI} \doi{10.1785/0120120010} \end{APACrefDOI}
\PrintBackRefs{\CurrentBib}

\bibitem [\protect \citeauthoryear {%
Helmstetter%
, Kagan%
\BCBL {}\ \BBA {} Jackson%
}{%
Helmstetter%
\ \protect \BOthers {.}}{%
{\protect \APACyear {2006}}%
}]{%
helmstetter2006comparison}
\APACinsertmetastar {%
helmstetter2006comparison}%
\begin{APACrefauthors}%
Helmstetter, A.%
, Kagan, Y\BPBI Y.%
\BCBL {}\ \BBA {} Jackson, D\BPBI D.%
\end{APACrefauthors}%
\unskip\
\newblock
\APACrefYearMonthDay{2006}{}{}.
\newblock
{\BBOQ}\APACrefatitle {Comparison of short-term and time-independent earthquake
  forecast models for southern California} {Comparison of short-term and
  time-independent earthquake forecast models for southern california}.{\BBCQ}
\newblock
\APACjournalVolNumPages{Bull. Seismol. Soc. Am.}{96}{1}{90--106}.
\PrintBackRefs{\CurrentBib}

\bibitem [\protect \citeauthoryear {%
Hoyt%
\ \BBA {} Schatten%
}{%
Hoyt%
\ \BBA {} Schatten%
}{%
{\protect \APACyear {1998}}%
}]{%
hoyt1998group}
\APACinsertmetastar {%
hoyt1998group}%
\begin{APACrefauthors}%
Hoyt, D\BPBI V.%
\BCBT {}\ \BBA {} Schatten, K\BPBI H.%
\end{APACrefauthors}%
\unskip\
\newblock
\APACrefYearMonthDay{1998}{}{}.
\newblock
{\BBOQ}\APACrefatitle {Group sunspot numbers: A new solar activity
  reconstruction} {Group sunspot numbers: A new solar activity
  reconstruction}.{\BBCQ}
\newblock
\APACjournalVolNumPages{Sol. Phys}{179}{1}{189--219}.
\PrintBackRefs{\CurrentBib}

\bibitem [\protect \citeauthoryear {%
Huc%
\ \BBA {} Main%
}{%
Huc%
\ \BBA {} Main%
}{%
{\protect \APACyear {2003}}%
}]{%
Huc2003}
\APACinsertmetastar {%
Huc2003}%
\begin{APACrefauthors}%
Huc, M.%
\BCBT {}\ \BBA {} Main, I\BPBI G.%
\end{APACrefauthors}%
\unskip\
\newblock
\APACrefYearMonthDay{2003}{}{}.
\newblock
{\BBOQ}\APACrefatitle {{Anomalous stress diffusion in earthquake triggering:
  Correlation length, time dependence, and directionality}} {{Anomalous stress
  diffusion in earthquake triggering: Correlation length, time dependence, and
  directionality}}.{\BBCQ}
\newblock
\APACjournalVolNumPages{J. Geophys. Res. Solid Earth}{108}{B7}{}.
\newblock
\begin{APACrefDOI} \doi{10.1029/2001jb001645} \end{APACrefDOI}
\PrintBackRefs{\CurrentBib}

\bibitem [\protect \citeauthoryear {%
Hutchinson%
, England%
, Santoso%
\BCBL {}\ \BBA {} Hogg%
}{%
Hutchinson%
\ \protect \BOthers {.}}{%
{\protect \APACyear {2013}}%
}]{%
Hutchinson2013}
\APACinsertmetastar {%
Hutchinson2013}%
\begin{APACrefauthors}%
Hutchinson, D\BPBI K.%
, England, M\BPBI H.%
, Santoso, A.%
\BCBL {}\ \BBA {} Hogg, A\BPBI M\BPBI C.%
\end{APACrefauthors}%
\unskip\
\newblock
\APACrefYearMonthDay{2013}{}{}.
\newblock
{\BBOQ}\APACrefatitle {{Interhemispheric asymmetry in transient global warming:
  The role of Drake Passage}} {{Interhemispheric asymmetry in transient global
  warming: The role of Drake Passage}}.{\BBCQ}
\newblock
\APACjournalVolNumPages{Geophys. Res. Lett.}{40}{8}{1587--1593}.
\newblock
\begin{APACrefDOI} \doi{10.1002/grl.50341} \end{APACrefDOI}
\PrintBackRefs{\CurrentBib}

\bibitem [\protect \citeauthoryear {%
Ide%
}{%
Ide%
}{%
{\protect \APACyear {2013}}%
}]{%
Ide2013}
\APACinsertmetastar {%
Ide2013}%
\begin{APACrefauthors}%
Ide, S.%
\end{APACrefauthors}%
\unskip\
\newblock
\APACrefYearMonthDay{2013}{}{}.
\newblock
{\BBOQ}\APACrefatitle {{The proportionality between relative plate velocity and
  seismicity in subduction zones}} {{The proportionality between relative plate
  velocity and seismicity in subduction zones}}.{\BBCQ}
\newblock
\APACjournalVolNumPages{Nat. Geosci.}{6}{9}{780--784}.
\newblock
\begin{APACrefDOI} \doi{10.1038/ngeo1901} \end{APACrefDOI}
\PrintBackRefs{\CurrentBib}

\bibitem [\protect \citeauthoryear {%
Jordan%
\ \protect \BOthers {.}}{%
Jordan%
\ \protect \BOthers {.}}{%
{\protect \APACyear {2011}}%
}]{%
jordan2011operational}
\APACinsertmetastar {%
jordan2011operational}%
\begin{APACrefauthors}%
Jordan, T\BPBI H.%
, Chen, Y\BHBI T.%
, Gasparini, P.%
, Madariaga, R.%
, Main, I.%
, Marzocchi, W.%
\BDBL {}Zschau, J.%
\end{APACrefauthors}%
\unskip\
\newblock
\APACrefYearMonthDay{2011}{}{}.
\newblock
{\BBOQ}\APACrefatitle {Operational earthquake forecasting. State of knowledge
  and guidelines for utilization} {Operational earthquake forecasting. state of
  knowledge and guidelines for utilization}.{\BBCQ}
\newblock
\APACjournalVolNumPages{Ann. Geophys.}{54}{4}{361--391}.
\PrintBackRefs{\CurrentBib}

\bibitem [\protect \citeauthoryear {%
Kagan%
}{%
Kagan%
}{%
{\protect \APACyear {2004}}%
}]{%
Kagan2004}
\APACinsertmetastar {%
Kagan2004}%
\begin{APACrefauthors}%
Kagan, Y\BPBI Y.%
\end{APACrefauthors}%
\unskip\
\newblock
\APACrefYearMonthDay{2004}{}{}.
\newblock
{\BBOQ}\APACrefatitle {{Short-term properties of earthquake catalogs and models
  of earthquake source}} {{Short-term properties of earthquake catalogs and
  models of earthquake source}}.{\BBCQ}
\newblock
\APACjournalVolNumPages{Bull. Seismol. Soc. Am.}{94}{4}{1207--1228}.
\newblock
\begin{APACrefDOI} \doi{10.1785/012003098} \end{APACrefDOI}
\PrintBackRefs{\CurrentBib}

\bibitem [\protect \citeauthoryear {%
King%
}{%
King%
}{%
{\protect \APACyear {1996}}%
}]{%
King1996}
\APACinsertmetastar {%
King1996}%
\begin{APACrefauthors}%
King, T.%
\end{APACrefauthors}%
\unskip\
\newblock
\APACrefYearMonthDay{1996}{}{}.
\newblock
{\BBOQ}\APACrefatitle {{Quantifying nonlinearity and geometry in time series of
  climate}} {{Quantifying nonlinearity and geometry in time series of
  climate}}.{\BBCQ}
\newblock
\APACjournalVolNumPages{Quat. Sci. Rev.}{15}{4}{247--266}.
\newblock
\begin{APACrefDOI} \doi{10.1016/0277-3791(95)00060-7} \end{APACrefDOI}
\PrintBackRefs{\CurrentBib}

\bibitem [\protect \citeauthoryear {%
Lennartz%
, Livina%
, Bunde%
\BCBL {}\ \BBA {} Havlin%
}{%
Lennartz%
\ \protect \BOthers {.}}{%
{\protect \APACyear {2008}}%
}]{%
Lennartz2008}
\APACinsertmetastar {%
Lennartz2008}%
\begin{APACrefauthors}%
Lennartz, S.%
, Livina, V\BPBI N.%
, Bunde, A.%
\BCBL {}\ \BBA {} Havlin, S.%
\end{APACrefauthors}%
\unskip\
\newblock
\APACrefYearMonthDay{2008}{}{}.
\newblock
{\BBOQ}\APACrefatitle {{Long-term memory in earthquakes and the distribution of
  interoccurrence times}} {{Long-term memory in earthquakes and the
  distribution of interoccurrence times}}.{\BBCQ}
\newblock
\APACjournalVolNumPages{EPL}{81}{6}{3--7}.
\newblock
\begin{APACrefDOI} \doi{10.1209/0295-5075/81/69001} \end{APACrefDOI}
\PrintBackRefs{\CurrentBib}

\bibitem [\protect \citeauthoryear {%
Lippiello%
, {De Arcangelis}%
\BCBL {}\ \BBA {} Godano%
}{%
Lippiello%
\ \protect \BOthers {.}}{%
{\protect \APACyear {2008}}%
}]{%
Lippiello2008}
\APACinsertmetastar {%
Lippiello2008}%
\begin{APACrefauthors}%
Lippiello, E.%
, {De Arcangelis}, L.%
\BCBL {}\ \BBA {} Godano, C.%
\end{APACrefauthors}%
\unskip\
\newblock
\APACrefYearMonthDay{2008}{}{}.
\newblock
{\BBOQ}\APACrefatitle {{Influence of time and space correlations on earthquake
  magnitude}} {{Influence of time and space correlations on earthquake
  magnitude}}.{\BBCQ}
\newblock
\APACjournalVolNumPages{Phys. Rev. Lett.}{100}{3}{1--4}.
\newblock
\begin{APACrefDOI} \doi{10.1103/PhysRevLett.100.038501} \end{APACrefDOI}
\PrintBackRefs{\CurrentBib}

\bibitem [\protect \citeauthoryear {%
Lippiello%
, {De Arcangelis}%
\BCBL {}\ \BBA {} Godano%
}{%
Lippiello%
\ \protect \BOthers {.}}{%
{\protect \APACyear {2009}}%
}]{%
Lippiello2009a}
\APACinsertmetastar {%
Lippiello2009a}%
\begin{APACrefauthors}%
Lippiello, E.%
, {De Arcangelis}, L.%
\BCBL {}\ \BBA {} Godano, C.%
\end{APACrefauthors}%
\unskip\
\newblock
\APACrefYearMonthDay{2009}{}{}.
\newblock
{\BBOQ}\APACrefatitle {{Role of static stress diffusion in the spatiotemporal
  organization of aftershocks}} {{Role of static stress diffusion in the
  spatiotemporal organization of aftershocks}}.{\BBCQ}
\newblock
\APACjournalVolNumPages{Phys. Rev. Lett.}{103}{3}{}.
\newblock
\begin{APACrefDOI} \doi{10.1103/PhysRevLett.103.038501} \end{APACrefDOI}
\PrintBackRefs{\CurrentBib}

\bibitem [\protect \citeauthoryear {%
Lippiello%
, Godano%
\BCBL {}\ \BBA {} {De Arcangelis}%
}{%
Lippiello%
\ \protect \BOthers {.}}{%
{\protect \APACyear {2012}}%
}]{%
Lippiello2012}
\APACinsertmetastar {%
Lippiello2012}%
\begin{APACrefauthors}%
Lippiello, E.%
, Godano, C.%
\BCBL {}\ \BBA {} {De Arcangelis}, L.%
\end{APACrefauthors}%
\unskip\
\newblock
\APACrefYearMonthDay{2012}{}{}.
\newblock
{\BBOQ}\APACrefatitle {{The earthquake magnitude is influenced by previous
  seismicity}} {{The earthquake magnitude is influenced by previous
  seismicity}}.{\BBCQ}
\newblock
\APACjournalVolNumPages{Geophys. Res. Lett.}{39}{5}{}.
\newblock
\begin{APACrefDOI} \doi{10.1029/2012GL051083} \end{APACrefDOI}
\PrintBackRefs{\CurrentBib}

\bibitem [\protect \citeauthoryear {%
Livina%
\ \protect \BOthers {.}}{%
Livina%
\ \protect \BOthers {.}}{%
{\protect \APACyear {2003}}%
}]{%
Livina2003a}
\APACinsertmetastar {%
Livina2003a}%
\begin{APACrefauthors}%
Livina, V\BPBI N.%
, Ashkenazy, Y.%
, Braun, P.%
, Monetti, R.%
, Bunde, A.%
\BCBL {}\ \BBA {} Havlin, S.%
\end{APACrefauthors}%
\unskip\
\newblock
\APACrefYearMonthDay{2003}{}{}.
\newblock
{\BBOQ}\APACrefatitle {{Nonlinear volatility of river flux fluctuations}}
  {{Nonlinear volatility of river flux fluctuations}}.{\BBCQ}
\newblock
\APACjournalVolNumPages{Phys. Rev. E}{67}{4}{4}.
\newblock
\begin{APACrefDOI} \doi{10.1103/PhysRevE.67.042101} \end{APACrefDOI}
\PrintBackRefs{\CurrentBib}

\bibitem [\protect \citeauthoryear {%
Livina%
, Havlin%
\BCBL {}\ \BBA {} Bunde%
}{%
Livina%
\ \protect \BOthers {.}}{%
{\protect \APACyear {2005}}%
}]{%
Livina2005}
\APACinsertmetastar {%
Livina2005}%
\begin{APACrefauthors}%
Livina, V\BPBI N.%
, Havlin, S.%
\BCBL {}\ \BBA {} Bunde, A.%
\end{APACrefauthors}%
\unskip\
\newblock
\APACrefYearMonthDay{2005}{}{}.
\newblock
{\BBOQ}\APACrefatitle {{Memory in the Occurrence of Earthquakes}} {{Memory in
  the Occurrence of Earthquakes}}.{\BBCQ}
\newblock
\APACjournalVolNumPages{Phys. Rev. Lett.}{95}{20}{208501}.
\newblock
\begin{APACrefDOI} \doi{10.1103/PhysRevLett.95.208501} \end{APACrefDOI}
\PrintBackRefs{\CurrentBib}

\bibitem [\protect \citeauthoryear {%
Lombardi%
}{%
Lombardi%
}{%
{\protect \APACyear {2015}}%
}]{%
Lombardi2015}
\APACinsertmetastar {%
Lombardi2015}%
\begin{APACrefauthors}%
Lombardi, A\BPBI M.%
\end{APACrefauthors}%
\unskip\
\newblock
\APACrefYearMonthDay{2015}{}{}.
\newblock
{\BBOQ}\APACrefatitle {{Estimation of the parameters of ETAS models by
  Simulated Annealing}} {{Estimation of the parameters of ETAS models by
  Simulated Annealing}}.{\BBCQ}
\newblock
\APACjournalVolNumPages{Sci. Rep.}{5}{1}{8417}.
\newblock
\begin{APACrefDOI} \doi{10.1038/srep08417} \end{APACrefDOI}
\PrintBackRefs{\CurrentBib}

\bibitem [\protect \citeauthoryear {%
Marsan%
\ \BBA {} Lenglin{\'{e}}%
}{%
Marsan%
\ \BBA {} Lenglin{\'{e}}%
}{%
{\protect \APACyear {2008}}%
}]{%
Marsan2008}
\APACinsertmetastar {%
Marsan2008}%
\begin{APACrefauthors}%
Marsan, D.%
\BCBT {}\ \BBA {} Lenglin{\'{e}}, O.%
\end{APACrefauthors}%
\unskip\
\newblock
\APACrefYearMonthDay{2008}{}{}.
\newblock
{\BBOQ}\APACrefatitle {{Extending earthquakes' reach through cascading}}
  {{Extending earthquakes' reach through cascading}}.{\BBCQ}
\newblock
\APACjournalVolNumPages{Science}{319}{5866}{1076--1079}.
\newblock
\begin{APACrefDOI} \doi{10.1126/science.1148783} \end{APACrefDOI}
\PrintBackRefs{\CurrentBib}

\bibitem [\protect \citeauthoryear {%
Marzocchi%
\ \BBA {} Lombardi%
}{%
Marzocchi%
\ \BBA {} Lombardi%
}{%
{\protect \APACyear {2009}}%
}]{%
Marzocchi2009}
\APACinsertmetastar {%
Marzocchi2009}%
\begin{APACrefauthors}%
Marzocchi, W.%
\BCBT {}\ \BBA {} Lombardi, A\BPBI M.%
\end{APACrefauthors}%
\unskip\
\newblock
\APACrefYearMonthDay{2009}{}{}.
\newblock
{\BBOQ}\APACrefatitle {{Real-time forecasting following a damaging earthquake}}
  {{Real-time forecasting following a damaging earthquake}}.{\BBCQ}
\newblock
\APACjournalVolNumPages{Geophys. Res. Lett.}{36}{21}{}.
\newblock
\begin{APACrefDOI} \doi{10.1029/2009GL040233} \end{APACrefDOI}
\PrintBackRefs{\CurrentBib}

\bibitem [\protect \citeauthoryear {%
Ogata%
}{%
Ogata%
}{%
{\protect \APACyear {1988}}%
}]{%
Ogata1988}
\APACinsertmetastar {%
Ogata1988}%
\begin{APACrefauthors}%
Ogata, Y.%
\end{APACrefauthors}%
\unskip\
\newblock
\APACrefYearMonthDay{1988}{}{}.
\newblock
{\BBOQ}\APACrefatitle {{Statistical Models for Earthquake Occurrences and
  Residual Analysis for Point Processes}} {{Statistical Models for Earthquake
  Occurrences and Residual Analysis for Point Processes}}.{\BBCQ}
\newblock
\APACjournalVolNumPages{J. Am. Stat. Assoc.}{83}{401}{9--27}.
\newblock
\begin{APACrefDOI} \doi{10.1080/01621459.1988.10478560} \end{APACrefDOI}
\PrintBackRefs{\CurrentBib}

\bibitem [\protect \citeauthoryear {%
Ogata%
}{%
Ogata%
}{%
{\protect \APACyear {1998}}%
}]{%
Ogata1998}
\APACinsertmetastar {%
Ogata1998}%
\begin{APACrefauthors}%
Ogata, Y.%
\end{APACrefauthors}%
\unskip\
\newblock
\APACrefYearMonthDay{1998}{}{}.
\newblock
{\BBOQ}\APACrefatitle {{Space-Time Point-Process Models for Earthquake
  Occurrences}} {{Space-Time Point-Process Models for Earthquake
  Occurrences}}.{\BBCQ}
\newblock
\APACjournalVolNumPages{Ann. Inst. Stat. Math.}{50}{2}{379--402}.
\newblock
\begin{APACrefDOI} \doi{10.1023/A:1003403601725} \end{APACrefDOI}
\PrintBackRefs{\CurrentBib}

\bibitem [\protect \citeauthoryear {%
Petrillo%
\ \BBA {} Lippiello%
}{%
Petrillo%
\ \BBA {} Lippiello%
}{%
{\protect \APACyear {2021}}%
}]{%
petrillo2021testing}
\APACinsertmetastar {%
petrillo2021testing}%
\begin{APACrefauthors}%
Petrillo, G.%
\BCBT {}\ \BBA {} Lippiello, E.%
\end{APACrefauthors}%
\unskip\
\newblock
\APACrefYearMonthDay{2021}{}{}.
\newblock
{\BBOQ}\APACrefatitle {Testing of the foreshock hypothesis within an epidemic
  like description of seismicity} {Testing of the foreshock hypothesis within
  an epidemic like description of seismicity}.{\BBCQ}
\newblock
\APACjournalVolNumPages{Geophys. J. Int.}{225}{2}{1236--1257}.
\PrintBackRefs{\CurrentBib}

\bibitem [\protect \citeauthoryear {%
Richards-Dinger%
, Stein%
\BCBL {}\ \BBA {} Toda%
}{%
Richards-Dinger%
\ \protect \BOthers {.}}{%
{\protect \APACyear {2010}}%
}]{%
Felzer}
\APACinsertmetastar {%
Felzer}%
\begin{APACrefauthors}%
Richards-Dinger, K.%
, Stein, R\BPBI S.%
\BCBL {}\ \BBA {} Toda, S.%
\end{APACrefauthors}%
\unskip\
\newblock
\APACrefYearMonthDay{2010}{}{}.
\newblock
{\BBOQ}\APACrefatitle {{Decay of aftershock density with distance does not
  indicate triggering by dynamic stress}} {{Decay of aftershock density with
  distance does not indicate triggering by dynamic stress}}.{\BBCQ}
\newblock
\APACjournalVolNumPages{Nature}{467}{7315}{583--586}.
\newblock
\begin{APACrefDOI} \doi{10.1038/nature09402} \end{APACrefDOI}
\PrintBackRefs{\CurrentBib}

\bibitem [\protect \citeauthoryear {%
Saichev%
\ \BBA {} Sornette%
}{%
Saichev%
\ \BBA {} Sornette%
}{%
{\protect \APACyear {2006}}%
}]{%
Saichev2006}
\APACinsertmetastar {%
Saichev2006}%
\begin{APACrefauthors}%
Saichev, A.%
\BCBT {}\ \BBA {} Sornette, D.%
\end{APACrefauthors}%
\unskip\
\newblock
\APACrefYearMonthDay{2006}{}{}.
\newblock
{\BBOQ}\APACrefatitle {{“Universal” Distribution of Interearthquake Times
  Explained}} {{“Universal” Distribution of Interearthquake Times
  Explained}}.{\BBCQ}
\newblock
\APACjournalVolNumPages{Phys. Rev. Lett.}{97}{7}{078501}.
\newblock
\begin{APACrefDOI} \doi{10.1103/PhysRevLett.97.078501} \end{APACrefDOI}
\PrintBackRefs{\CurrentBib}

\bibitem [\protect \citeauthoryear {%
Schreiber%
\ \BBA {} Schmitz%
}{%
Schreiber%
\ \BBA {} Schmitz%
}{%
{\protect \APACyear {1996}}%
}]{%
Schreiber1996}
\APACinsertmetastar {%
Schreiber1996}%
\begin{APACrefauthors}%
Schreiber, T.%
\BCBT {}\ \BBA {} Schmitz, A.%
\end{APACrefauthors}%
\unskip\
\newblock
\APACrefYearMonthDay{1996}{}{}.
\newblock
{\BBOQ}\APACrefatitle {{Improved surrogate data for nonlinearity tests}}
  {{Improved surrogate data for nonlinearity tests}}.{\BBCQ}
\newblock
\APACjournalVolNumPages{Phys. Rev. Lett.}{77}{4}{635--638}.
\newblock
\begin{APACrefDOI} \doi{10.1103/PhysRevLett.77.635} \end{APACrefDOI}
\PrintBackRefs{\CurrentBib}

\bibitem [\protect \citeauthoryear {%
Seif%
, Mignan%
, Zechar%
, Werner%
\BCBL {}\ \BBA {} Wiemer%
}{%
Seif%
\ \protect \BOthers {.}}{%
{\protect \APACyear {2017}}%
}]{%
Seif2017}
\APACinsertmetastar {%
Seif2017}%
\begin{APACrefauthors}%
Seif, S.%
, Mignan, A.%
, Zechar, J\BPBI D.%
, Werner, M\BPBI J.%
\BCBL {}\ \BBA {} Wiemer, S.%
\end{APACrefauthors}%
\unskip\
\newblock
\APACrefYearMonthDay{2017}{}{}.
\newblock
{\BBOQ}\APACrefatitle {{Estimating ETAS: The effects of truncation, missing
  data, and model assumptions}} {{Estimating ETAS: The effects of truncation,
  missing data, and model assumptions}}.{\BBCQ}
\newblock
\APACjournalVolNumPages{J. Geophys. Res. Solid Earth}{122}{1}{449--469}.
\newblock
\begin{APACrefDOI} \doi{10.1002/2016JB012809} \end{APACrefDOI}
\PrintBackRefs{\CurrentBib}

\bibitem [\protect \citeauthoryear {%
Sornette%
, Utkin%
\BCBL {}\ \BBA {} Saichev%
}{%
Sornette%
\ \protect \BOthers {.}}{%
{\protect \APACyear {2008}}%
}]{%
Sornette2008}
\APACinsertmetastar {%
Sornette2008}%
\begin{APACrefauthors}%
Sornette, D.%
, Utkin, S.%
\BCBL {}\ \BBA {} Saichev, A.%
\end{APACrefauthors}%
\unskip\
\newblock
\APACrefYearMonthDay{2008}{}{}.
\newblock
{\BBOQ}\APACrefatitle {{Solution of the nonlinear theory and tests of
  earthquake recurrence times}} {{Solution of the nonlinear theory and tests of
  earthquake recurrence times}}.{\BBCQ}
\newblock
\APACjournalVolNumPages{Phys. Rev. E}{77}{6}{1--10}.
\newblock
\begin{APACrefDOI} \doi{10.1103/PhysRevE.77.066109} \end{APACrefDOI}
\PrintBackRefs{\CurrentBib}

\bibitem [\protect \citeauthoryear {%
Touati%
, Naylor%
\BCBL {}\ \BBA {} Main%
}{%
Touati%
\ \protect \BOthers {.}}{%
{\protect \APACyear {2009}}%
}]{%
Touati2009}
\APACinsertmetastar {%
Touati2009}%
\begin{APACrefauthors}%
Touati, S.%
, Naylor, M.%
\BCBL {}\ \BBA {} Main, I\BPBI G.%
\end{APACrefauthors}%
\unskip\
\newblock
\APACrefYearMonthDay{2009}{}{}.
\newblock
{\BBOQ}\APACrefatitle {{Origin and Nonuniversality of the Earthquake Interevent
  Time Distribution}} {{Origin and Nonuniversality of the Earthquake Interevent
  Time Distribution}}.{\BBCQ}
\newblock
\APACjournalVolNumPages{Phys. Rev. Lett.}{102}{16}{168501}.
\newblock
\begin{APACrefDOI} \doi{10.1103/PhysRevLett.102.168501} \end{APACrefDOI}
\PrintBackRefs{\CurrentBib}

\bibitem [\protect \citeauthoryear {%
Utsu%
}{%
Utsu%
}{%
{\protect \APACyear {1961}}%
}]{%
Utsu1961}
\APACinsertmetastar {%
Utsu1961}%
\begin{APACrefauthors}%
Utsu, T.%
\end{APACrefauthors}%
\unskip\
\newblock
\APACrefYearMonthDay{1961}{}{}.
\newblock
{\BBOQ}\APACrefatitle {{A statistical study on the occurrence of af-
  tershocks}} {{A statistical study on the occurrence of af-
  tershocks}}.{\BBCQ}
\newblock
\APACjournalVolNumPages{Geophys. Mag.}{30}{}{521--605}.
\PrintBackRefs{\CurrentBib}

\bibitem [\protect \citeauthoryear {%
Utsu%
}{%
Utsu%
}{%
{\protect \APACyear {1972}}%
}]{%
UTSU1972}
\APACinsertmetastar {%
UTSU1972}%
\begin{APACrefauthors}%
Utsu, T.%
\end{APACrefauthors}%
\unskip\
\newblock
\APACrefYearMonthDay{1972}{}{}.
\newblock
{\BBOQ}\APACrefatitle {{Aftershocks and Earthquake Statistics (3) : Analyses of
  the Distribution of Earthquakes in Magnitude, Time and Space with Special
  Consideration to Clustering Characteristics of Earthquake Occurrence(1)}}
  {{Aftershocks and Earthquake Statistics (3) : Analyses of the Distribution of
  Earthquakes in Magnitude, Time and Space with Special Consideration to
  Clustering Characteristics of Earthquake Occurrence(1)}}.{\BBCQ}
\newblock
\APACjournalVolNumPages{J. Fac. Sci. Hokkaido Univ. Ser. 7,
  Geophys.}{4}{1}{1--42}.
\PrintBackRefs{\CurrentBib}

\bibitem [\protect \citeauthoryear {%
Woessner%
, Christophersen%
, {Douglas Zechar}%
\BCBL {}\ \BBA {} Monelli%
}{%
Woessner%
\ \protect \BOthers {.}}{%
{\protect \APACyear {2010}}%
}]{%
Woessner2010}
\APACinsertmetastar {%
Woessner2010}%
\begin{APACrefauthors}%
Woessner, J.%
, Christophersen, A.%
, {Douglas Zechar}, J.%
\BCBL {}\ \BBA {} Monelli, D.%
\end{APACrefauthors}%
\unskip\
\newblock
\APACrefYearMonthDay{2010}{}{}.
\newblock
{\BBOQ}\APACrefatitle {{Building self-consistent, short-term earthquake
  probability (STEP) models: Improved strategies and calibration procedures}}
  {{Building self-consistent, short-term earthquake probability (STEP) models:
  Improved strategies and calibration procedures}}.{\BBCQ}
\newblock
\APACjournalVolNumPages{Ann. Geophys.}{53}{3}{141--154}.
\newblock
\begin{APACrefDOI} \doi{10.4401/ag-4812} \end{APACrefDOI}
\PrintBackRefs{\CurrentBib}

\bibitem [\protect \citeauthoryear {%
Yano%
, Takeda%
, Matsubara%
\BCBL {}\ \BBA {} Shiomi%
}{%
Yano%
\ \protect \BOthers {.}}{%
{\protect \APACyear {2017}}%
}]{%
Yano2017}
\APACinsertmetastar {%
Yano2017}%
\begin{APACrefauthors}%
Yano, T\BPBI E.%
, Takeda, T.%
, Matsubara, M.%
\BCBL {}\ \BBA {} Shiomi, K.%
\end{APACrefauthors}%
\unskip\
\newblock
\APACrefYearMonthDay{2017}{}{}.
\newblock
{\BBOQ}\APACrefatitle {{Japan unified hIgh-resolution relocated catalog for
  earthquakes (JUICE): Crustal seismicity beneath the Japanese Islands}}
  {{Japan unified hIgh-resolution relocated catalog for earthquakes (JUICE):
  Crustal seismicity beneath the Japanese Islands}}.{\BBCQ}
\newblock
\APACjournalVolNumPages{Tectonophysics}{702}{}{19--28}.
\newblock
\begin{APACrefDOI} \doi{10.1016/j.tecto.2017.02.017} \end{APACrefDOI}
\PrintBackRefs{\CurrentBib}

\bibitem [\protect \citeauthoryear {%
Zaliapin%
\ \BBA {} Ben-Zion%
}{%
Zaliapin%
\ \BBA {} Ben-Zion%
}{%
{\protect \APACyear {2013}}%
}]{%
Yehuda2013}
\APACinsertmetastar {%
Yehuda2013}%
\begin{APACrefauthors}%
Zaliapin, I.%
\BCBT {}\ \BBA {} Ben-Zion, Y.%
\end{APACrefauthors}%
\unskip\
\newblock
\APACrefYearMonthDay{2013}{}{}.
\newblock
{\BBOQ}\APACrefatitle {{Earthquake clusters in southern California I:
  Identification and stability}} {{Earthquake clusters in southern California
  I: Identification and stability}}.{\BBCQ}
\newblock
\APACjournalVolNumPages{J. Geophys. Res. Solid Earth}{118}{6}{2847--2864}.
\newblock
\begin{APACrefDOI} \doi{10.1002/jgrb.50179} \end{APACrefDOI}
\PrintBackRefs{\CurrentBib}

\bibitem [\protect \citeauthoryear {%
Zaliapin%
, Gabrielov%
, Keilis-Borok%
\BCBL {}\ \BBA {} Wong%
}{%
Zaliapin%
\ \protect \BOthers {.}}{%
{\protect \APACyear {2008}}%
}]{%
Zaliapin2008}
\APACinsertmetastar {%
Zaliapin2008}%
\begin{APACrefauthors}%
Zaliapin, I.%
, Gabrielov, A.%
, Keilis-Borok, V.%
\BCBL {}\ \BBA {} Wong, H.%
\end{APACrefauthors}%
\unskip\
\newblock
\APACrefYearMonthDay{2008}{}{}.
\newblock
{\BBOQ}\APACrefatitle {{Clustering analysis of seismicity and aftershock
  identification}} {{Clustering analysis of seismicity and aftershock
  identification}}.{\BBCQ}
\newblock
\APACjournalVolNumPages{Phys. Rev. Lett.}{101}{1}{4--7}.
\newblock
\begin{APACrefDOI} \doi{10.1103/PhysRevLett.101.018501} \end{APACrefDOI}
\PrintBackRefs{\CurrentBib}

\bibitem [\protect \citeauthoryear {%
Zhang%
\ \protect \BOthers {.}}{%
Zhang%
\ \protect \BOthers {.}}{%
{\protect \APACyear {2020}}%
}]{%
Zhang2019}
\APACinsertmetastar {%
Zhang2019}%
\begin{APACrefauthors}%
Zhang, Y.%
, Fan, J.%
, Marzocchi, W.%
, Shapira, A.%
, Hofstetter, R.%
, Havlin, S.%
\BCBL {}\ \BBA {} Ashkenazy, Y.%
\end{APACrefauthors}%
\unskip\
\newblock
\APACrefYearMonthDay{2020}{}{}.
\newblock
{\BBOQ}\APACrefatitle {{Scaling laws in earthquake memory for interevent times
  and distances}} {{Scaling laws in earthquake memory for interevent times and
  distances}}.{\BBCQ}
\newblock
\APACjournalVolNumPages{Phys. Rev. Res.}{2}{1}{013264}.
\newblock
\begin{APACrefDOI} \doi{10.1103/PhysRevResearch.2.013264} \end{APACrefDOI}
\PrintBackRefs{\CurrentBib}

\bibitem [\protect \citeauthoryear {%
Zhang%
\ \protect \BOthers {.}}{%
Zhang%
\ \protect \BOthers {.}}{%
{\protect \APACyear {2021}}%
}]{%
Zhang2020}
\APACinsertmetastar {%
Zhang2020}%
\begin{APACrefauthors}%
Zhang, Y.%
, Zhou, D.%
, Fan, J.%
, Marzocchi, W.%
, Ashkenazy, Y.%
\BCBL {}\ \BBA {} Havlin, S.%
\end{APACrefauthors}%
\unskip\
\newblock
\APACrefYearMonthDay{2021}{}{}.
\newblock
{\BBOQ}\APACrefatitle {{Improved earthquake aftershocks forecasting model based
  on long-term memory}} {{Improved earthquake aftershocks forecasting model
  based on long-term memory}}.{\BBCQ}
\newblock
\APACjournalVolNumPages{New J. Phys}{23}{}{042001}.
\newblock
\begin{APACrefDOI} \doi{10.1088/1367-2630/abeb46} \end{APACrefDOI}
\PrintBackRefs{\CurrentBib}

\bibitem [\protect \citeauthoryear {%
Zhuang%
}{%
Zhuang%
}{%
{\protect \APACyear {2012}}%
}]{%
Zhuang2012}
\APACinsertmetastar {%
Zhuang2012}%
\begin{APACrefauthors}%
Zhuang, J.%
\end{APACrefauthors}%
\unskip\
\newblock
\APACrefYearMonthDay{2012}{}{}.
\newblock
{\BBOQ}\APACrefatitle {{Long-term earthquake forecasts based on the
  epidemic-type aftershock sequence (ETAS) model for short-term clustering}}
  {{Long-term earthquake forecasts based on the epidemic-type aftershock
  sequence (ETAS) model for short-term clustering}}.{\BBCQ}
\newblock
\APACjournalVolNumPages{Res. Geophys.}{2}{1}{8}.
\newblock
\begin{APACrefDOI} \doi{10.4081/rg.2012.e8} \end{APACrefDOI}
\PrintBackRefs{\CurrentBib}

\bibitem [\protect \citeauthoryear {%
Zhuang%
, Murru%
, Falcone%
\BCBL {}\ \BBA {} Guo%
}{%
Zhuang%
\ \protect \BOthers {.}}{%
{\protect \APACyear {2019}}%
}]{%
Zhuang2019}
\APACinsertmetastar {%
Zhuang2019}%
\begin{APACrefauthors}%
Zhuang, J.%
, Murru, M.%
, Falcone, G.%
\BCBL {}\ \BBA {} Guo, Y.%
\end{APACrefauthors}%
\unskip\
\newblock
\APACrefYearMonthDay{2019}{}{}.
\newblock
{\BBOQ}\APACrefatitle {{An extensive study of clustering features of seismicity
  in Italy from 2005 to 2016}} {{An extensive study of clustering features of
  seismicity in Italy from 2005 to 2016}}.{\BBCQ}
\newblock
\APACjournalVolNumPages{Geophys. J. Int.}{216}{1}{302--318}.
\newblock
\begin{APACrefDOI} \doi{10.1093/gji/ggy428} \end{APACrefDOI}
\PrintBackRefs{\CurrentBib}

\bibitem [\protect \citeauthoryear {%
Zhuang%
, Werner%
, Harte%
, Hainzl%
\BCBL {}\ \BBA {} Zhou%
}{%
Zhuang%
\ \protect \BOthers {.}}{%
{\protect \APACyear {2010}}%
}]{%
Zhuang2010}
\APACinsertmetastar {%
Zhuang2010}%
\begin{APACrefauthors}%
Zhuang, J.%
, Werner, M\BPBI J.%
, Harte, D.%
, Hainzl, S.%
\BCBL {}\ \BBA {} Zhou, S.%
\end{APACrefauthors}%
\unskip\
\newblock
\APACrefYearMonthDay{2010}{}{}.
\newblock
{\BBOQ}\APACrefatitle {{Basic models of seismicity}} {{Basic models of
  seismicity}}.{\BBCQ}
\newblock
\APACjournalVolNumPages{Community Online Resour. Stat. Seism.
  Anal.}{}{}{2--41}.
\newblock
\begin{APACrefDOI} \doi{10.5078/corssa-47845067.} \end{APACrefDOI}
\PrintBackRefs{\CurrentBib}

\end{thebibliography}

%
%
%
%
%

\end{document}